\documentclass[a4paper,11pt]{article}
\usepackage[margin=1in]{geometry}
\usepackage{amsmath,amssymb,amsfonts}
\usepackage{algorithmic}
\usepackage{graphicx,color,xcolor}
\usepackage{textcomp}
\pdfoutput=1
\usepackage{microtype}
\usepackage{graphicx}
\usepackage{subfigure}
\usepackage{booktabs} 

\usepackage{framed}
\usepackage{xcolor}
\usepackage{tcolorbox}
\newcommand{\ra}[1]{\renewcommand{\arraystretch}{#1}}
\usepackage{tablefootnote}

\usepackage{hyperref}
\usepackage{amsmath,amsthm,amssymb,amsfonts}
\usepackage{bm}
\usepackage{pifont}
\usepackage{stackengine}

\usepackage{algorithm}
\usepackage{algorithmic}
\usepackage{enumerate}
\usepackage{bbm}

\theoremstyle{plain}
\newtheorem{theorem}{Theorem}[]
\newtheorem{corollary}{Corollary}[theorem]
\newtheorem{lemma}[]{Lemma}

\newtheorem{assumption}{Assumption}[]

\def\x{{\mathbf x}}
\def\z{{\mathbf z}}

\def\g{{\mathbf g}}
\def\h{{\mathbf h}}

\def\v{{\mathbf v}}

\def\y{{\mathbf y}}

\def\R{{\mathbb{R}}}

\def\E{{\mathcal E}}

\def\O{{\mathcal O}}

\begin{document}
\title{Accelerated Distributed Stochastic Non-Convex Optimization over Time-Varying Directed Networks}
\date{}
\author{Yiyue Chen, Abolfazl Hashemi,  Haris Vikalo
\thanks{This work was supported in part by NSF grants 1809327 and CNS-2313109.}
\thanks{Yiyue Chen and Haris Vikalo are with the Department of Electrical and Computer Engineering, 
University of Texas at Austin, Austin, TX 78712 USA. Abolfazl Hashemi is with the School of Electrical and Computer Engineering, Purdue University, West Lafayette, IN 47907, USA.}
\thanks{A preliminary version of this article is presented at the 2023 International Conference on Acoustics, Speech, and Signal Processing (ICASSP) \cite{chen2023accelerated}.}
}

\maketitle
	
\begin{abstract}
Distributed stochastic non-convex optimization problems have recently received attention due to the growing interest of signal processing, computer vision, and natural language processing communities in applications deployed over distributed learning systems (e.g., federated learning). We study the setting where the data is distributed across the nodes of a time-varying directed network, a topology suitable for modeling dynamic networks experiencing communication delays and straggler effects. The network nodes, which can access only their local objectives and query a stochastic first-order oracle to obtain gradient estimates, collaborate to minimize a global objective function by exchanging messages with their neighbors. We propose an algorithm, novel to this setting, that leverages stochastic gradient descent with momentum and gradient tracking to solve distributed non-convex optimization problems over time-varying networks. To analyze the algorithm, we tackle the challenges that arise when analyzing dynamic network systems which communicate gradient acceleration components. We prove that the algorithm's oracle complexity is $\mathcal{O}(1/\epsilon^{1.5})$, and that under Polyak-$\L$ojasiewicz condition the algorithm converges linearly to a steady error state. The proposed scheme is tested on several learning tasks: a non-convex logistic regression experiment on the MNIST dataset, an image classification task on the CIFAR-10 dataset, and an NLP classification test on the IMDB dataset. We further present numerical simulations with an objective that satisfies the PL condition. The results demonstrate superior performance of the proposed framework compared to the existing related methods.
\end{abstract}
\section{Introduction}

We study distributed non-convex optimization problems encountered in a variety of applications in machine learning (ML), signal processing, and control \cite{kempe2003gossip, ren2005consensus, nedic2009distributed}. Distributed learning frameworks aim to address limitations of centralized methods including the potentially high cost of communicating data to a central location, privacy and latency concerns, and data storage constraints \cite{assran2019stochastic}.
We model a distributed computing system via a time-varying directed network $\mathcal{G}(t) = (\mathcal{V}, \E(t))$, where $\mathcal{V} = \{1, \cdots, n \}$ denotes the set of $n$ nodes and $\E(t)$ is the collection of directed edges $(i, j) $, $i, j \in \mathcal{V}$, connecting the nodes at time $t$. In particular, if $(i, j) \in \E(t)$, there exists an edge from node $i$ to node $j$, and thus node $i$ can send messages to node $j$ at time $t$. Node $i$ has access only to its local data and the local loss function. 
The goal of the nodes in the network is to collaboratively minimize a global loss function, i.e., solve
\begin{equation}\label{eq:prob}
\min_{\mathbf{x} \in \R^d} \left[f(\mathbf{x}):=\frac{1}{n}\sum_{i=1}^n f_i(\mathbf{x})\right],
\end{equation}
where $f_i(\mathbf{x}): \mathbb{R}^d \to \mathbb{R}$ for $i \in [n]:=\left\{1, ..., n \right\}$ denotes the non-convex objective that the device at node $i$ minimizes locally; in the machine learning context, this describes the setting where each node trains a local model by optimizing a cost function with $d$ parameters and collaborates with other nodes to find the global model.

In a departure from the existing work focused on \textit{undirected networks} \cite{hashemi2021benefits, xin2020decentralized, xin2021hybrid, pu2018distributed}, we study distributed non-convex optimization over \textit{time-varying directed networks} where each node minimizes its local objective utilizing a stochastic gradient obtained by querying a local stochastic first-order oracle, i.e., the nodes use noisy estimates of the local gradient at the query point. Unlike undirected networks, directed communication topologies take into account a number of practical considerations including asymmetry in the communication links (e.g., in the multi-agent applications) and the straggler effects stemming from imposing synchronized communication. Furthermore, time-varying networks characterize the communication link delay or failure in real-world applications. 

\begin{table*}[t]
\caption{A comparison of algorithms for decentralized optimization over directed graphs. SFO and IFO stand for stochastic and incremental first-order oracles, respectively. SGP relies on a bounded dissimilarity assumption to obtain the stated result. GT-HSGD utilizes the non-standard doubly-stochastic requirement for the weight matrix. The stated results for SGP and GT-HSGD are semi-asymptotic, meaning they require a large number of iterations to achieve the stated convergence bounds. DS(double-stochasticity) is satisfied by all undirected networks and a small collection of directed networks; DN(dynamic network) refers to time-varying networks, i.e., indicates that the network is changing over time.
}
\centering
\ra{1}
\scriptsize 
\begin{tabular*}{.85\linewidth}{@{}ccccc@{}}\toprule
Algorithm & DS & DN & Oracle Complexity  & Remarks \\ \midrule
\begin{tabular}{@{}c@{}} SGP/Push-SGD  \cite{assran2019stochastic} \end{tabular}  & No & Yes &  $\O(\frac{1}{n\epsilon^2})$ & \begin{tabular}{@{}c@{}} SFO, bounded dissimilarity  \\ only for large $T$\end{tabular}
\\\midrule 
\begin{tabular}{@{}c@{}} Push-DIGing  \cite{nedic2017achieving}  \end{tabular} & No & Yes & $\O(\ln \frac{1}{\epsilon})$ & \begin{tabular}{@{}c@{}} deterministic \\ Strong convexity, smoothness \end{tabular} \\\midrule
\begin{tabular}{@{}c@{}} Di-CS-SVRG  \cite{chen2021communication} \end{tabular}  & No & Yes &$\O(\ln \frac{1}{\epsilon})$ & \begin{tabular}{@{}c@{}} IFO, strong convexity \\ smoothness \end{tabular} \\\midrule
\begin{tabular}{@{}c@{}} Push-SAGA  \cite{qureshi2021push} \end{tabular}  & No & No & $\O(\ln \frac{1}{\epsilon})$ & \begin{tabular}{@{}c@{}} IFO, strong convexity \\ smoothness \end{tabular}\\\midrule
\begin{tabular}{@{}c@{}} S-ADDOPT  \cite{qureshi2020s} \end{tabular}  & No & No & $\O( \frac{1}{\epsilon})$ & \begin{tabular}{@{}c@{}} SFO, strong convexity \\ smoothness \end{tabular}\\\midrule
\begin{tabular}{@{}c@{}} GT-HSGD  \cite{xin2021hybrid} \end{tabular}  & Yes & No & $\O( \frac{1}{n\epsilon^{1.5}})$ & \begin{tabular}{@{}c@{}} SFO, mean-squared smoothness \\ doubly stochastic weight matrix \\ only for large $T$\end{tabular} \\\midrule
{\bf Push-ASGD (This work)} & No & Yes & $\O( \frac{1}{\epsilon^{1.5}})$ & \begin{tabular}{@{}c@{}} SFO, mean-squared smoothness  \end{tabular} \\\midrule
{\bf Push-ASGD (This work)} & No & Yes &$\O( \ln \frac{1}{\epsilon})$ & \begin{tabular}{@{}c@{}} SFO, mean-squared smoothness,\\PL condition  \end{tabular} \\
\bottomrule
\end{tabular*}
\label{tb:comp}
\end{table*}
\vspace{-2mm}
\subsection{Related work and significance}

Decentralized non-convex optimization problems with stochastic first-order oracles have been extensively studied in the context of undirected networks, where doubly stochastic weight matrices lead to convergence guarantees. Those studies include the decentralized stochastic gradient descent (DSGD) and its variants, \cite{kar2012distributed, lian2017can, hashemi2021benefits}, which combine decentralized average consensus with local gradient updates. Although DSGD is often effective and relatively simple to implement, it is unstable in settings that involve heterogeneous data  \cite{xin2020decentralized}. This has motivated the search for more robust schemes which combine decentralized bias-correction techniques, gradient tracking, and primal-dual methods \cite{xu2015augmented, nedic2017achieving, pu2018distributed, das2020faster}. Building on top of such techniques, GT-HSGD \cite{xin2021hybrid} leverages SARAH-type variance reduction schemes (see, e.g., \cite{cutkosky2019momentum, pham2020proxsarah}) to further reduce oracle complexities under the so-called mean-squared smoothness \cite{fang2018spider}. However, GT-HSGD still relies on doubly stochastic weight matrices hindering its practical feasibility in applications involving asymmetric communication.

While there has been extensive prior work on distributed optimization over the family of networks characterized by doubly-stochastic mixing matrix, e.g., undirected networks and some special cases of directed networks, practical systems experience transmission failures and/or bandwidth limitations which imply asymmetric or uni-directional communication between network nodes. In such scenarios, more general directed graphs that do not satisfy doubly-stochastic property are a better-suited network model \cite{assran2019stochastic}; however, the design of convergent algorithms for distributed optimization over general directed graphs brings forth new challenges. In particular, we recall that to ensure convergence of the algorithms for decentralized optimization over undirected networks, the weight (mixing) matrix $W_m$ should be symmetric and doubly stochastic. Indeed, mixing matrix characterizes communication over a network: when doubly stochastic, the decentralized algorithm reaches the average consensus model since $\lim_{T \to \infty}\Pi_{t=1}^T W_m = \frac{ \mathbf{1} \mathbf{1}^T}{n}$. When a network is directed, however, the communication links are asymmetric and the corresponding mixing matrix is generally not doubly stochastic. There, it holds that $\lim_{T \to \infty}\Pi_{t=1}^T W_m = \mathbf{\pi} \mathbf{1}^T$ but unlike the undirected network scenario $\pi_i \neq \frac{1}{n}$, which implies convergence to a biased weighted average of local models. As a remedy, distributed optimization schemes for directed networks often deploy auxiliary variables to help deal with the communication asymmetry. For instance, frequently used subgradient-push algorithm \cite{kempe2003gossip,nedic2014distributed} and its variants \cite{qureshi2020s}, which operates on column-stochastic mixing matrices, introduces local normalization scalars to de-bias the weighted average and thus ensure convergence. In another line of related work, \cite{xi2017distributed,chen2021decentralized} introduce auxiliary variables of the same dimension as the local model parameters to keep track of the local parameter variations and avoid division (a nonlinear operation) deployed by the subgradient-push algorithm. When the objective is smooth and strongly convex, a linear rate can be achieved by using constant step size  \cite{nedic2017achieving, saadatniaki2018optimization, chen2021communication, nedich2022ab}. Further acceleration of the convergence rate can be achieved adopting momentum-based methods \cite{nguyen2023accelerated}.

Distributed, stochastic {\em non-convex} optimization over directed time-varying graphs has received relatively little attention \cite{assran2019stochastic}. Even though some of the above optimization algorithms (developed with convex objectives in mind) can be applied to distributed non-convex optimization problems, they either converge at a slow rate or apply without any theoretical guarantees. Aiming to achieve robust and provably fast performance, we present and analyze an algorithm that relies on gradient-push, global gradient tracking, and a local hybrid gradient estimator with momentum to solve distributed, stochastic non-convex optimization problems that arise in modern distributed ML tasks. Our contributions are summarized as follows: 

   1.~We study (to our knowledge, previously not pursued in literature) the problem of distributed non-convex optimization under a stochastic first-order oracle (SFO) over directed time-varying networks. We propose for this setting novel variance-reduced algorithm, Push-ASGD, which leverages gradient-push, global gradient tracking, and hybrid gradient estimation methods. In particular, we devise a new stochastic gradient estimator used locally by each node, and design a framework that operates with column stochastic weight matrices.

    2.~To analyze convergence of Push-ASGD, we address challenges brought forward by the exchange of gradient acceleration components across directed time-varying networks. We prove that, under the mean-squared smoothness, the algorithm attains an $\epsilon$-accurate first-order stationary solution with an oracle complexity of $\O(1/\epsilon^{1.5})$. To our knowledge, this is the first algorithm to come with such guarantees in the context of stochastic nonconvex optimization over directed networks. We further show that for objective functions satisfying Polyak-$\L$ojasiewicz (PL) condition, when using constant step size the algorithm converges linearly to a steady state with small error.  
    
    3.~We validate the proposed algorithm on various distributed learning problems including image classification and natural language processing via deep learning, demonstrating its superior accuracy/convergence compared to relevant existing techniques. We also test its performance in simulations involving an objective function that satisfies the PL condition. The results demonstrate superior accuracy/convergence of Push-ASGD compared to the relevant benchmarking algorithms.

\vspace{-2mm}
\section{Preliminaries}



Assume that $n$ nodes are collaboratively solving the decentralized non-convex optimization problem \eqref{eq:prob} while communicating over a time-varying directed network.
Each node $i$ in the network trains a local model and computes a sequence of local model estimates $\{\x_t^i \}$ towards a first-order stationary point of the global objective $f$, starting from a pre-specified initialization point $\x_0^i$. To update its local model estimate, node $i$ accesses a random local data point $ \mathbf{\xi}^i_{t} $ and queries stochastic first-order oracle for a stochastic gradient, $\nabla f_i(\mathbf{x}, \mathbf{\xi}^i_{t} )$, given the input $\x$. To characterize the sequence of random local data points, we consider the filtration induced by the data points of all the nodes in the network,
\begin{equation}
\begin{aligned}
    \mathcal{F}_0  = \{\Omega, \phi \},\;\;
    \mathcal{F}_t  = \sigma(\{\mathbf{\xi}^i_{0}, \mathbf{\xi}^i_{1}, \cdots, \mathbf{\xi}^i_{t-1}, i \in \mathcal{V} \}), \forall t \geq 1,
\end{aligned}
\end{equation}
where $\phi$ is the empty set and $\{\mathcal{F}_t\}$ is an increasing family of $\sigma$-algebras. The input vector $\x$ at iteration $t$ is $\mathcal{F}_t$-measurable. We denote the probability space by $\{\Omega, \mathbb{P}, \mathcal{F} \}$. We make the following assumptions on the stochastic first-order oracles, the global objective, and the communication network.

\begin{assumption}\label{assumption1}
For all $i \in \mathcal{V}$ and all $t \geq 0$, we assume:
1.~Unbiasedness of the conditional expectation
    \begin{align*}
        \mathbb{E}[\nabla f_i(\mathbf{x}, \mathbf{\xi}^i_{t} ) |\mathcal{F}_t ] = \nabla f_i(\x).
    \end{align*}
2.~Bounded variance of the estimated gradient,
    \begin{align*}
        \mathbb{E}\|\nabla f_i(\mathbf{x}, \mathbf{\xi}^i_{t} ) - \nabla f_i(x) \|^2 \leq \nu_i^2.
    \end{align*}
It will be convenient to introduce $\Bar{\nu}^2 = \frac{1}{n}\sum_{i=1}^n \nu_i^2$.\\
3.~Independent random selection, i.e., the random vectors $\{\mathbf{\xi}^i_{0}, \mathbf{\xi}^i_{1}, \cdots, \mathbf{\xi}^i_{t-1}\}$, $i \in \mathcal{V}$, are independent.\\
4.~The mean-squared smoothness,
    \begin{align*}
        \mathbb{E}\|\nabla f_i(\mathbf{x}, \mathbf{\xi}^i_{t} ) - \nabla f_i(\mathbf{y}, \mathbf{\xi}^i_{t} ) \|^2 \leq L^2 \mathbb{E}\|\x - \mathbf{y} \|^2.
    \end{align*}
\end{assumption}
The first three assumptions are widely used in the analysis of stochastic first-order optimization algorithms \cite{bottou2010large}. The last assumption requires $L$-smoothness of the stochastic gradient on average with respect to any two inputs. The mean-squared smoothness further implies smoothness of each local objective $f_i$, and consequently implies $L$-smoothness of the global objective \cite{fang2018spider,arjevani2019lower}. 

\begin{assumption}\label{assumption2}
The global objective is lower bounded, 
\begin{equation*}
    f^* = \inf_{\x } f(\x) > -\infty.
\end{equation*}
\end{assumption}

\begin{assumption}\label{assumption3}
The network is directed and time-varying. At time $t$, the network is strongly-connected with a column stochastic weight matrix $W_m^{(t)}$. 
\end{assumption}

The above assumption on network topology is standard in distributed optimization \cite{nedic2017achieving} and more general than the static network assumption in \cite{assran2019stochastic}. Elaborating on Assumption~3, let us consider the mixing matrix $W_m^{(t)} := [w_{ij}^{(t)}]$ which captures properties of the communication links in the network;
$w_{ij}^{(t)} = \frac{1}{d_j^{out, t}+1}>0 $ if and only if 
$ (j, i) \in \E(t)$ 
or $i = j$, where $d_j^{out, t} $ is the out-degree of agent $j$ at time $t$. 
We assume that node $i$ knows which nodes it sends messages to, i.e., the $i$-th column of $W_m^{(t)}$ are known to node $i$. 
The column stochastic mixing matrix $W_m^{(t)}$ has the left eigenvector $\mathbf{1}_n$ and a positive right eigenvector 
$\pi^{(t)}$, i.e., $\mathbf{1}_n^\top W_m^{(t)} = \mathbf{1}_n^\top $ and $W_m^{(t)}\pi^{(t)} = \pi^{(t)}$.



\vspace{-2mm}

\section{The Push-ASGD Algorithm}
\setlength{\intextsep}{0pt}
\definecolor{light-gray}{gray}{0.9}
\begin{algorithm}[t]
\caption{Push Accelerated Stochastic Gradient Descent Algorithm (Push-ASGD)}
\begin{algorithmic}[1]
   \STATE {\bfseries Input:} Initialize $\mathbf{x}^i_0 = \Bar{\mathbf{x}}_0$; $y_0^i = 1$ ; $\mathbf{z}^i_0 = \mathbf{x}^i_0 $; step size $\alpha$; $\beta \in (0, 1)$; time-varying column-stochastic mixing matrix $W_m^{(t)} := [w^{(t)}_{ij}]$; $T \in \mathbf{Z}^+$\;
\STATE Sample $b$ local data points $\{ \mathbf{\xi}_{0, r}^i\}_{r=1}^b$ and initialize the gradient $\mathbf{g}_0^i = \frac{1}{b} \sum_{r=1}^b\nabla f_i(\mathbf{z}^i_0, \mathbf{\xi}_{0, r}^i)$ and the gradient estimator $\mathbf{v}^i_{0} = \mathbf{g}_0^i$; \;
   \FOR{$t=0$ {\bfseries to} $T-1$}
   \STATE Update the local estimate of the solution
   
   \colorbox{light-gray}{$\mathbf{x}_{t+1}^i = \sum_{j=1}^n w^{(t)}_{ij}(\mathbf{x}_t^j - \alpha \mathbf{g}_{t}^j) $}
   \STATE Update auxiliary variables
   
    \colorbox{light-gray}{$y_{t+1}^i = \sum_{j=1}^n w^{(t)}_{ij}y_t^j , \ \mathbf{z}_{t+1}^i =\frac{ \mathbf{x}_{t+1}^i} { y_{t+1}^i }$} 
    \STATE
    Sample $\mathbf{\xi}_t^i$ and update the local stochastic gradient estimator
    
    \colorbox{light-gray}{$ \mathbf{v}^i_{t+1} =  \nabla f_i(\mathbf{z}^i_{t+1}, \mathbf{\xi}^i_{t}) + (1-\beta)(\mathbf{v}^i_{t} -\nabla f_i(\mathbf{z}^i_{t}, \mathbf{\xi}^i_{t} ) ) $} \;
    \STATE
    Update the gradient tracker 
    
    \colorbox{light-gray}{$ \mathbf{g}_{t+1}^i = \sum_{j=1}^n w^{(t)}_{ij}(\mathbf{g}_t^j + \mathbf{v}_{t+1}^j - \mathbf{v}_{t}^j ) $} \;
   \ENDFOR
\end{algorithmic}
\label{non_cvx_algorithm}
\end{algorithm}

In this section we present an algorithm for distributed non-convex optimization over directed time-varying networks. At a high level, the algorithm relies on the push-sum protocol \cite{kempe2003gossip} to perform average consensus, and deploys a stochastic gradient estimator of the unknown global gradient while simultaneously reducing the variance/noise in the local updates via momentum.

At the beginning, all the nodes use the same initial model $\Bar{\mathbf{x}}_0 $. At iteration $t$, node $i$ updates its local model $\x^i_t$ by fusing the messages $\mathbf{x}_t^j$ received from its neighbors according to
\begin{equation}
    \mathbf{x}_{t+1}^i = \sum_{j=1}^n w_{ij}^{(t)}(\mathbf{x}_t^j - \alpha \mathbf{g}_{t}^j),
\end{equation}
where $\mathbf{g}_{t}^j$ denotes the local stochastic gradient estimate specified below. Since $W_m^{(t)}$ is column-stochastic, the product of $W_m^{(t)}$ over a time duration $s$, $\Pi_{k = 1}^s W_m^{(t+k)} $, generally differ from $\frac{ \mathbf{1} \mathbf{1}^T}{n}$, biasing each node to a different model. Therefore, following \cite{nedic2016stochastic}, for each node $i$ with local model $\x^i_t$ at time $t$ we introduce an auxiliary scalar $y^i_t$ and compute a recovering model $\z^i_t = \x^i_t / y^i_t$, enabling de-biasing the local model fused by the mixing matrix. In other words, while the average of $\x^i_t$ is not preserved due to directed communication, the average of the de-biased quantities $\z^i_t$ will be preserved \cite{nedic2014distributed}.

In addition to addressing challenges that arise from directed communication, the proposed algorithm deals with two sources of variance that hamper the convergence: a local variance that stems from noise in the local stochastic gradients, and a global variance that stems from the heterogeneity of the nodes' data. To address the former, we rely on momentum-based variance reduction, while to tackle the latter we deploy gradient tracking.

Let $\nabla f_i(\mathbf{z}^i_{t+1}, \mathbf{\xi}^i_{t})$ and $\nabla f_i(\mathbf{z}^i_{t}, \mathbf{\xi}^i_{t})$ denote stochastic gradients obtained after querying the local stochastic first-order oracle with $\mathbf{z}^i_{t+1} $ and $\mathbf{z}^i_{t}$, respectively. The momentum-type update of gradient $\v^i_{t+1}$ is then found as
\begin{equation}
    \mathbf{v}^i_{t+1} =  \nabla f_i(\mathbf{z}^i_{t+1}, \mathbf{\xi}^i_{t}) + (1-\beta)(\mathbf{v}^i_{t} -\nabla f_i(\mathbf{z}^i_{t}, \mathbf{\xi}^i_{t} ) ),
    \label{eq:update_v}
\end{equation}
where $\beta$ denotes the momentum step size controlling the direction of the gradient adjustment term $\mathbf{v}^i_{t} -\nabla f_i(\mathbf{z}^i_{t}, \mathbf{\xi}^i_{t} ) $.
When $\beta = 1$, \eqref{eq:update_v} reduces to the vanilla stochastic gradient descent, while when $\beta = 0$, \eqref{eq:update_v} reduces to a SARAH-type gradient update \cite{xin2020near}; neither can achieve the same oracle complexity as the recursive estimator deploying $\beta \in (0,1)$. This recursive estimator can reduce the variance of the stochastic gradient estimates in both centralized and distributed optimization problems \cite{cutkosky2019momentum,sun2020improving}. 
Finally, the estimate of the global gradient is found via gradient tracking as
\begin{equation}
    \mathbf{g}_{t+1}^i = \sum_{j=1}^n w_{ij}^{(t)}(\mathbf{g}_t^j + \mathbf{v}_{t+1}^j - \mathbf{v}_{t}^j ) ,
\end{equation}
where the gradient information from neighboring nodes is used to ensure convergence to the first-order stationary point of the global objective.

The above procedure is formalized as Algorithm~1 and in the remainder of the paper referred to as the Push-ASGD (\underline{Push} \underline{A}ccelerated \underline{S}tochastic \underline{G}radient \underline{D}escent) algorithm.
\vspace{-2mm}
\section{Convergence Analysis}
We proceed by analytically showing that Push-ASGD achieves $\O(1/\epsilon^{1.5})$ SFO complexity, and that under the PL condition it linearly converges to a small steady-state error. Below, the first theorem establishes the $\O(1/\epsilon^{1.5})$ complexity while the second theorem establishes linear convergence of Push-ASGD. 



For convenience, we re-write the key terms in Algorithm \ref{non_cvx_algorithm} as
\begin{align*}
    y_{t+1} & = W^{(t)}_m y_t \\ 
    \z_{t+1, n \times d} & = \Tilde{W}_m^{(t)} (\z_{t, n \times d} - \alpha \h_{t, n \times d}) \\
    \h_{t+1, n \times d} & =  \Tilde{W}_m^{(t)} \h_{t, n \times d} + \Tilde{W}_m^{(t)} Y^{-1}_t (\v_{t+1, n \times d} - \v_{t, n \times d}),
\end{align*}
where $ \Tilde{W}_m^{(t)} = Y^{-1}_{t+1} W_m^{(t)} Y_t$, $Y_t = diag(y_t)$ and $\h_{t, n \times d} = Y^{-1}_{t} \g_{t, n \times d} $. Moreover, $ \z_{t, n \times d} = [(\z_t^1)^\top ; \cdots ; (\z_t^n)^\top ]$ (similar for $\g_{t, n \times d}, \v_{t, n \times d} $); note that $n \times d$ in the subscript indicates dimension of a matrix. Finally, $\Tilde{W}_m^{(t)}$ is a row-stochastic mixing matrix; there exists a stochastic vector sequence $\{\phi_t \}$ such that $\phi_{t+1}^T \Tilde{W}_m^{(t)} = \phi_t^T $.\footnote{Further details are in the proof of Lemma \ref{lemma1} in Appendix A.}

Before stating the theorem, it will be convenient to introduce the global vectors in $\R^{nd}$
\begin{equation}
\begin{aligned}
&\x_t = [(\x_t^1)^\top \cdots, (\x_t^n)^\top ]^\top, \;\;
\z_t = [(\z_t^1)^\top \cdots, (\z_t^n)^\top ]^\top, \\
&\g_t = [(\mathbf{g}_t^1)^\top \cdots, (\mathbf{g}_t^n)^\top ]^\top , \;\;\;\;
\v_t = [(\v_t^1)^\top \cdots, (\v_t^n)^\top ]^\top, \\
&\nabla \mathbf{f}(\z_{t}) = [\nabla f_1(\z_t^1)^\top \cdots \nabla f_n(\z_t^n) ]^\top, \;\; \h_t = [(\h_t^1)^\top \cdots (\h_t^n)^\top ]^\top,
\end{aligned}
\end{equation}
as well as the averaged global vectors in $\R^{nd}$,
\begin{equation}
    \begin{aligned}
    \bar{\mathbf{x}}_t & = \frac{1}{n} [(\sum_j \x_t^j)^\top \cdots (\sum_j \x_t^j)^\top]^\top, \\
    \bar{\mathbf{v}}_t & = \frac{1}{n} [(\sum_j \v_t^j)^\top \cdots (\sum_j \v_t^j)^\top]^\top, \\
    \hat{\z}_t & = [ (\sum_j [\phi_t]_j \z_t^j)^\top \cdots (\sum_j [\phi_t]_j \z_t^j)^\top ], \\
    \nabla \bar{\mathbf{f}}(\mathbf{z_t}) & = \frac{1}{n} [(\sum_j \nabla f_j(\z_t^j))^\top \cdots (\sum_j \nabla f_j(\z_t^j))^\top]^\top, \\
\end{aligned}
\end{equation}
and the global time-varying matrices
\begin{equation}
    \begin{aligned}
    W^{(t)}  = W_m^{(t)} \otimes I_d, \ \Tilde{W}^{(t)} = \Tilde{W}_m^{(t)} \otimes I_d.
    \end{aligned}
\end{equation}
The updates of global vectors $\z_t, \h_t \in \R^{nd} $ can be written as
\begin{align*}
    \z_{t+1} & = \Tilde{W}^{(t)} (\z_t - \alpha \h_t), \\
    \h_{t+1} & =  \Tilde{W}^{(t)} \h_t + \Tilde{W}^{(t)} (Y^{-1}_t \otimes I_d ) (\v_{t+1} - \v_t).
\end{align*}
For the global vectors, let the $\mathfrak{L} $-norm be defined as
$\mathfrak{L}^2(\z_t, \phi_t) = \|(diag(\phi_t)^{\frac{1}{2}} \otimes I_d )(\z_t - \hat{\z}_t ) \|_F^2 $, where $\hat{\z}_t$ is the $\phi_t$-weighted average of $\z_t$. $\mathfrak{L}$-norm is induced by a sequence of time-varying stochastic vectors and facilitates the derivation of one-step consensus contraction addressed in Lemma \ref{lemma1} and \ref{lemma2}.

\begin{theorem}\label{thm:main}
Suppose Assumptions \ref{assumption1} -- \ref{assumption3} hold. Let step size $\alpha$ satisfy
\begin{equation}
\begin{aligned}
    & 0 < \alpha \leq \min \{ \frac{(1-\delta^2)^2}{48\delta^2\|Y^{-1} \| \sqrt{ (2 L^2 \phi_m (n+1) +  8 L^4 \phi_m (n+1))}}, \\
    & \frac{1}{2L}, \frac{1-\delta^2}{8\delta \|Y^{-1} \|^{1/2} \sqrt[\frac{1}{4}]{ (6L^2 \phi_m(n+1) +L^2 \phi_m) [\frac{18  L^2 }{(1-\delta^2)}  +36 L^4  ]}} \}.
    \end{aligned}
\end{equation}
Moreover, let the momentum step size $\beta$ be such that
\begin{equation}
      48 L^2 \alpha^2 \leq \beta<1.
\end{equation}
Then it holds that
\begin{equation}\label{eq:main-bound}
\begin{aligned}
    & \frac{1}{T}\sum_{t=0}^{T-1}\mathbb{E} \|\nabla f(\bar{\mathbf{x}}_{t}) \|^2 \\
    & \leq \frac{2\mathbb{E} (f(\bar{\mathbf{x}}_{0}) - f(\bar{\mathbf{x}}_{T})) }{\alpha T } + \frac{2}{\beta T} \mathbb{E}[\|\bar{ \mathbf{v}}_0 - \nabla \bar{\mathbf{f}}(\mathbf{z_0}) \|^2] + 4\beta \bar{\nu}^2    \\
    & + \frac{16\delta^2 \alpha^2}{(1-\delta^2)^4} (\frac{48 L^2\phi_m(n+1)}{\beta n^2} +\frac{4L^2\phi_m (n+1)}{n})   
    \\
    & \{ \frac{(1-\delta^2)}{T} \mathbb{E}\mathfrak{L}^2(\h_0, \phi_0) + 48\delta^2  \|Y^{-1} \|^2  \beta^2 n \bar{\nu}^2   \\ 
    & + 48\delta^2  \|Y^{-1} \|^2 L^2  [\frac{\beta}{T} \mathbb{E}[\| \mathbf{v}_0 - \nabla \mathbf{f}(\mathbf{z_0}) \|^2 ] + 2\beta^3 n \bar{\nu}^2     ] 
    \}, \\
\end{aligned}
\end{equation}
where $ \delta  = \max_t \sqrt{1 - \frac{\min_i([\phi_{t+1}]_i)}{\max_i([\phi_t]_i) (n-1)^2 n^{2(n+2)}} } \in (0, 1)$ is the network contraction parameter; 
$\phi_m = d/\min_{t, i} [\phi_t]_i$ is proportional to the inverse of the smallest entry in $\{\phi_t\}$, stochastic vectors associated with time-varying mixing matrices; and 
$\|Y^{-1} \| = \sup_{t} \|Y_t^{-1} \| $.\footnote{The existence/bounds for $\delta$, $\phi_m$, and $\|Y^{-1} \|$ are discussed in the proofs of Lemma \ref{lemma1} and \ref{lemma2}.}

\end{theorem}
Essentially, the bound in \eqref{eq:main-bound} is specified by the initial function value, the initial gradient estimation error, and the variance of the gradient estimator. 

If step sizes $\alpha$ and $\beta$ are chosen as in the statement of the above theorem, the average gradient error converges to a steady-state error,
\begin{equation}
\begin{aligned}
    & \lim_{T \to \infty} \frac{1}{T} \sum_{t=0}^{T-1} \mathbb{E}\|\nabla f(\bar{\mathbf{x}}_{t}) \|^2 \\ & \leq \frac{2\mathbb{E} (f(\bar{\mathbf{x}}_{0}) - f(\bar{\mathbf{x}}_{T})) }{\alpha T } + \frac{2}{\beta T} \mathbb{E}[\|\bar{ \mathbf{v}}_0 - \nabla \bar{\mathbf{f}}(\mathbf{z_0}) \|^2] + 4\beta \bar{\nu}^2    \\
    & + \frac{768\delta^2 \alpha^2}{(1-\delta^2)^4} (\frac{48 L^2\phi_m(n+1)}{\beta n^2} +\frac{4L^2\phi_m (n+1)}{n})   
    \\ &  (\delta^2  \|Y^{-1} \|^2  \beta^2 n \bar{\nu}^2   + 2\delta^2  \|Y^{-1} \|^2 L^2  \beta^3 n \bar{\nu}^2   
    ).
\end{aligned}
\end{equation}

A closer inspection of the right-hand side reveals that by selecting appropriate $\alpha$ and $\beta$, one can provide non-asymptotic convergence guarantees. Appropriate choices of $\alpha$ and $\beta$ and the corresponding convergence rate are specified in the following corollary.

\colorbox{light-gray}{\parbox{0.9\columnwidth}
{
\begin{corollary}
There exist values of the parameters 
$\alpha = \O(\frac{1}{n^{1/2}T^{1/3}})$, $\beta = \O( \frac{1}{T^{2/3}})$ and $b = \O(\frac{T^{1/3}}{n})$ such that
\begin{equation}
    \frac{1}{T} \sum_{t=0}^{T-1} \mathbb{E}\|\nabla f(\bar{\mathbf{x}}_{t}) \|^2 \leq \O(\frac{1}{T^{2/3}}).
\end{equation}
\end{corollary}
}}


{\color{black}{Recall that in each iteration of Push-ASGD, each node in the network samples one local data point and queries the stochastic first-order oracle for the gradient. Given the parameters in the corollary above, Push-ASGD can reach an $\epsilon$-accurate stationary point of the global objective with the overall oracle complexity of $O(\frac{1}{\epsilon^{1.5}})$.

When comparing the oracle complexity of Push-ASGD with those of other SFO algorithms in Table \ref{tb:comp}, we note that Push-ASGD has more desirable complexity for large $T$ when $n$ is fixed; although GT-HSGD has better oracle complexity, that algorithm does not apply to general directed time-varying graphs.

Note that Theorem~\ref{thm:main} applies to any strongly connected network. In simulations we observe that the number of steps needed to reach desirable accuracy level is smaller in more densely connected networks because fewer gradient queries are required and more information is exchanged.

}}

Next, we introduce the PL condition and investigate convergence under this additional assumption.
\begin{assumption}\label{assumption4} The objective function satisfies the PL condition with parameter $\mu$,
\begin{equation}
\begin{aligned}
    \frac{1}{2} \|\nabla f(x) \|^2 \geq \mu (f(x)- f^*),
\end{aligned}
\end{equation}
where $f^*$ is the optimal value of the objective function.
\end{assumption}
\begin{theorem}\label{thm:main2}
Suppose Assumptions \ref{assumption1} -- \ref{assumption4} hold. Let the step size $\alpha$ satisfy 
\begin{equation}\label{alpha2}
\begin{aligned}
    0&  < \alpha \leq \min \{ \frac{1}{2L}, \frac{1-\delta^2}{\mu}, \\
    & \frac{(1-\delta^2)^4 \mu}{36864 \delta^4 \|Y^{-1} \|^2 L^2 \phi_m (n+1) (3-\delta^2) }, \\
    & \frac{ 1-\delta^2}{24\delta^2 L \|Y^{-1} \|\sqrt{(72 \phi_m (n + 1) \frac{1 }{(1-\delta^2)^2}   + 16 \phi_m (n + 1) \frac{ 1}{1-\delta^2})  }} \}. 
    \end{aligned}
\end{equation}
Moreover, let the momentum step size $\beta$ satisfy
\begin{equation}\label{beta2}
    \begin{aligned}
    & \max \{\frac{\alpha \mu}{2}, \frac{768\alpha}{ \mu }  [\frac{3 L^2 \delta^4 \phi_m (356 + 212n )\|Y^{-1} \|^2 }{(1-\delta^2)^4} + \frac{3L^2 }{2 } ] \\ &  + 768\alpha^2  [ \frac{2736 L^2 \delta^4 \phi_m( n + 1 ) \|Y^{-1} \|^2}{(1-\delta^2)^4}  +  \frac{3L^2  }{4(1-\delta^2)^2}]  \} \leq \beta < 1.
    \end{aligned}
\end{equation}
Then
$\mathbb{E}[f(\bar{\mathbf{x}}_{t+1}) - f^*] $ decays linearly at the rate of $\O((1-\frac{\alpha \mu}{4})^t)$ to a steady-state error, i.e.,
\begin{equation}
\begin{aligned}
    & \lim_{t \to \infty} \sup \mathbb{E}[f(\bar{\mathbf{x}}_{t+1}) - f^*]  \\
   & \leq (\frac{\alpha^2 L^2 }{4} + \frac{3 L^2 \alpha^2}{2\beta } ) (4\frac{C_{exp}}{\alpha} + 6 \beta^2 \bar{\nu}^2 + \frac{(1-\delta^2)^2}{\alpha } 3\beta^2\bar{\nu}^2 ) \\
   & + \frac{288  \beta \bar{\nu}^2 }{ \mu (1-\delta^2)^2}  [\frac{8 L^2 \phi_m \alpha^2 \delta^4 (60 + 56\beta^2)(n+1) }{(1-\delta^2)^2}  \|Y^{-1} \|^2  \\
   &  +   (\frac{(1-\delta^2)^2}{4} + \frac{L^2 \|Y^{-1} \|^2 \alpha^2 \delta^4 \phi_m (96\beta^2 + 144 \delta^2 + 96\beta^2 n )}{(1-\delta^2)^2})  ],
\end{aligned}
\end{equation}
where $C_{exp} = \max_{k_m } \{ (1-\frac{\alpha \mu}{2})[C^{k_m} \mathbf{u}_0]_4 + \frac{\alpha}{n}[C^{k_m} \mathbf{u}_0]_3 + \frac{\alpha L^2\phi_m(2n + 2)}{n}[C^{k_m} \mathbf{u}_0]_1 \} $ and \\
$\mathbf{u}_0 = [\mathbb{E} \mathfrak{L}^2(\z_0, \phi_0 ), \mathbb{E}\mathfrak{L}^2(\h_0, \phi_0), \mathbb{E}[\|\mathbf{v}_{0} - \nabla \mathbf{f}(\mathbf{z_{0}}) \|_F^2], \mathbb{E}[f(\bar{\mathbf{x}}_{0}) - f^*]]. $
\end{theorem}

Theorem \ref{thm:main2} implies that for small values of step sizes $\alpha$ and $\beta$ (which satisfy the above conditions), the steady-state error will be small. Moreover, the following corollary on the non-asymptotic convergence holds.

\colorbox{light-gray}{\parbox{0.9\columnwidth}
{
\begin{corollary}

Suppose Assumptions \ref{assumption1} - \ref{assumption4} and step size conditions (\ref{alpha2}) and (\ref{beta2}) are satisfied.
If the values of the step sizes are such that
$\alpha = \O(T^{-1})$ and $\beta = o(T^{-1})$, i.e., $\alpha \to 0$ faster than $\beta \to 0$, then non-asymptotic convergence is guaranteed,
\begin{equation}
     \lim_{t \to \infty} \sup \mathbb{E}[f(\bar{\mathbf{x}}_{t+1}) - f^*]  \to 0.
\end{equation}
\end{corollary}
}}
\vspace{-2mm}
\subsection{Sketch of the proof}
Here we briefly go over the main steps of the proofs of Theorems~\ref{thm:main}-\ref{thm:main2}; full details are presented in the appendix. First, we identify the main error terms that contribute to the overall convergence error of Push-ASGD. These include:

    1.~$\mathbb{E}[\mathfrak{L}^2(\z_t, \phi_t)]$: the consensus error that quantifies how far the local models are from their average formed via the weight matrix.
    
    2.~$\mathbb{E}\| \mathbf{v}_t - \nabla \mathbf{f}(\mathbf{z_t}) \|^2 $: error of the momentum-based stochastic gradient estimator.
    
    3.~$\mathbb{E}[\mathfrak{L}^2(\h_t, \phi_t)]$: error of the global gradient tracking estimator.
    
    4.~$\mathbb{E}\|\bar{ \mathbf{v}}_t - \nabla \bar{\mathbf{f}}(\mathbf{z_t}) \|^2 $: variance of the momentum-based stochastic gradient estimator.
    
    5.~$\mathbb{E}[f(\bar{\mathbf{x}}_{t+1}) - f^*] $: optimality gap, measuring the distance from the optimal function value.

Our aim is to derive recursive inequalities that relate these error terms to each other. Then, to prove Theorem \ref{thm:main} we derive upper bounds on the relevant terms for each iteration $t$ and sum them over $t$ from $0$ to $T$. Finally, we combine the intermediate steps to achieve the main result, i.e., establish a bound on the average gradient norm accumulated over the iterations. The major challenge in the analysis is to establish relationships between the following terms:

1.~the consensus errors of the time-varying directed network system;
    
2.~the combination of the global gradient tracking error originating due to communication of gradient information over the network, the stochastic gradient computation and the momentum term for convergence acceleration at local clients.



To start, we introduce a lemma specifying an upper bound on the consensus error at time $t$.
\begin{lemma}\label{lemma1}
Suppose Assumptions \ref{assumption1} -- \ref{assumption3} hold. Based on the updates of Push-ASGD,
\begin{equation}
\begin{aligned}
    \mathbb{E}[\mathfrak{L}^2(\z_{t+1}, \phi_{t+1})]
    & \leq \frac{1+\delta^2}{2}  \mathbb{E}\mathfrak{L}^2(\z_t, \phi_t) + \frac{2\delta^2 \alpha^2}{1-\delta^2}  \mathbb{E}\mathfrak{L}^2(\h_t, \phi_t),
\end{aligned}
\end{equation}
for some network topology parameter $0 < \delta < 1$ indicated in Theorem \ref{thm:main}.
\end{lemma}

Next, we present a lemma stating an upper bound on the gradient tracking error at time $t$. 
\begin{lemma}\label{lemma2}
Suppose Assumptions \ref{assumption1} -- \ref{assumption3} hold. Then the gradient tracking error satisfies
\begin{equation}
    \begin{aligned}
    \mathbb{E} [\mathfrak{L}^2(\h_{t+1}, \phi_{t+1})] 
    & \leq \frac{1+\delta^2}{2} \mathbb{E}[ \mathfrak{L}^2(\h_t, \phi_t)] \\
    & + \frac{8\delta^2 }{1-\delta^2} \|Y^{-1} \|^2 [ 3L^2[3\alpha^2 \mathbb{E}[\|\bar{\mathbf{v}}_{t} \|^2] \\ &  + 6\phi_m(n+1)\mathbb{E}[ \mathfrak{L}^2(\z_{t+1}, \phi_{t+1}) + \mathfrak{L}^2(\z_t, \phi_t) ]] \\
    &+ 3\beta^2 \mathbb{E}[\| \mathbf{v}_t - \nabla \mathbf{f}(\mathbf{z_t}) \|^2 ] + 3\beta^2 \bar{\nu}^2 n  ].
    \end{aligned}
\end{equation}
\end{lemma}

The following lemma states an upper bound on the error of the momentum-based stochastic gradient estimator.
\begin{lemma}\label{lemma3}
Suppose Assumptions \ref{assumption1} -- \ref{assumption3} hold. The error of the momentum-based stochastic gradient estimator satisfies
\begin{equation}
    \begin{aligned}
    & \mathbb{E}[\| \mathbf{v}_t - \nabla \mathbf{f}(\mathbf{z_t}) \|^2 ] 
     \leq (1 - \beta)^2 \mathbb{E}[\|\mathbf{v}_{t-1} - \nabla \mathbf{f}(\mathbf{z_{t-1}}) \|^2] + 2\beta^2 n \bar{\nu}^2 \\
    & + 6(1-\beta)^2L^2 [\alpha^2 \mathbb{E}[\|\bar{\mathbf{v}}_{t-1} \|^2] \\
    & + 2 \phi_m (n+1) \mathbb{E}[\mathfrak{L}^2(\z_t, \phi_t) + \mathfrak{L}^2(\z_{t-1}, \phi_{t-1}) \|^2 ]] \\
    \end{aligned}
\end{equation}
while for its averaged version holds that
\begin{equation}
    \begin{aligned}
    & \mathbb{E}[\|\bar{ \mathbf{v}}_t - \nabla \bar{\mathbf{f}}(\mathbf{z_t}) \|^2 ] 
     \leq (1 - \beta)^2 \mathbb{E}[\|\bar{ \mathbf{v}}_{t-1} - \nabla \bar{\mathbf{f}}(\mathbf{z_{t-1}}) \|^2] + \frac{2\beta^2  \bar{\nu}^2}{n} \\ & + \frac{6(1-\beta)^2L^2  }{n^2} [\alpha^2 \mathbb{E}[\|\bar{\mathbf{v}}_{t-1} \|^2] \\
    & + 2\phi_m(n+1) \mathbb{E} [\mathfrak{L}^2(\z_t, \phi_t) + \mathfrak{L}^2(\z_{t-1}, \phi_{t-1}) \|^2]].
    \end{aligned}
\end{equation}
\end{lemma}
Using the lemmas above, we proceed by conducting telescoping from $t = 0$ to $T$ for each of the four errors to arrive at an upper bound for the sum of the errors. This leads to the following lemma for the squared gradient bound.
\begin{lemma}\label{lemma4}
The accumulated expected gradient norm at $\bar{\mathbf{x}}_t$ satisfies
\begin{equation}\label{eq:proof-sketch-final}
    \begin{aligned}
    & \sum_{t=0}^{T-1}\mathbb{E} \|\nabla [f(\bar{\mathbf{x}}_{t}) \|^2] 
    \leq \frac{2 \mathbb{E}(f(\bar{\mathbf{x}}_{0}) - f(\bar{\mathbf{x}}_{T})) }{\alpha } -\frac{1}{2}\sum_{t=0}^{T-1}\mathbb{E}\|\bar{\mathbf{v}}_{t} \|^2 \\ & + 2\sum_{t=0}^{T-1} \mathbb{E} \|\bar{\mathbf{v}}_t - \nabla \bar{\mathbf{f}}(\mathbf{z_t}) \|^2  +\frac{4L^2\phi_m(n+1)}{n}\sum_{t=0}^{T-1}\mathbb{E}[\mathfrak{L}^2(\z_t, \phi_t )].
    \end{aligned}
\end{equation}
\end{lemma}
To complete the proof of Theorem \ref{thm:main} we use the upper bounds on the last three sums on the right-hand-side of (\ref{eq:proof-sketch-final}), 
canceling out the sum $\sum_{t=0}^{T-1}\mathbb{E}\|\bar{\mathbf{v}}_{t} \|^2 $ which appears with a negative sign in \eqref{eq:proof-sketch-final}. Doing so requires imposing limitations on $\alpha$ and $\beta$, the learning rates of Push-ASGD, as stated in Theorem \ref{thm:main}.

Next, we outline the proof of Theorem \ref{thm:main2}, starting by building a linear inequality system on top of the error bound lemmas via incorporating the PL condition. In particular, one first needs to show that the largest eigenvalue of the coefficient matrix for the linear inequality system is strictly less than $1$, guaranteeing linear convergence. Then, one needs to find an upper bound on the time-varying residual terms, including $\mathbb{E}\|\bar{\mathbf{v}}_{t} \|^2$, that holds for all $t$. The proof can finally be complected by imposing conditions on the step sizes $\alpha$ and $\beta$. The main challenges in this analysis stem from the following:
\begin{enumerate}
\item due to non-doubly-stochastic mixing matrices, the error terms in the linear inequality systems are considerably more involved than in the case of undirected static networks;
\item the dynamic residual term $\mathbf{r}_t$ contains $\mathbb{E}\|\bar{\mathbf{v}}_{t} \|^2$ and its upper bound.
\end{enumerate}

To proceed, we use the PL condition to establish the following recursive relationship.
\begin{lemma}\label{lemma5}
Suppose Assumptions \ref{assumption1} - \ref{assumption4} hold and the step size $\alpha$ is such that $\alpha \leq \frac{1}{2L}$. Then
\begin{equation}
    \begin{aligned}
    & \mathbb{E}[f(\bar{\mathbf{x}}_{t+1}) - f^*] 
    \leq \mathbb{E}[(1-\frac{\alpha \mu}{2}) (f(\bar{\mathbf{x}}_t) - f^* ) - \frac{\alpha}{4} \|\bar{\mathbf{v}}_{t} \|^2 \\
    & + \frac{\alpha}{n} \|\mathbf{v}_t - \nabla \mathbf{f}(\mathbf{z_t}) \|^2 + \frac{ \alpha L^2 \phi_m (2n + 2)\mathfrak{L}^2(\z_t, \phi_t) }{n}  ].
    \end{aligned}
\end{equation}
\end{lemma}

\noindent
To form the linear inequality system, let us introduce
\begin{align*}
    \mathbf{u}_{k+1} = \begin{bmatrix} \mathbb{E}[\mathfrak{L}^2(\z_{k+1}, \phi_{k+1})] \\ \mathbb{E}[\mathfrak{L}^2 (\h_{k+1}, \phi_{k+1})] \\ \mathbb{E}[\|\mathbf{v}_{k+1} - \nabla \mathbf{f}(\mathbf{z_{k+1}}) \|_F^2] \\ \mathbb{E}[f(\bar{\mathbf{x}}_{k+1}) - f^*] \end{bmatrix},
\end{align*}
\begin{equation}\label{eq:c-def}
    \begin{aligned}
        C_1 & = [\frac{1+\delta^2}{2},  \frac{2\delta^2\alpha^2}{1-\delta^2},  0,  0] \\
        C_2 & = [\frac{144\delta^2 \|Y^{-1} \|^2L^2 \phi_m (2n + 2)}{1-\delta^2}, \frac{1+\delta^2}{2} \\ & + \frac{288 \delta^4 \|Y^{-1} \|^2L^2 \phi_m \alpha^2 (n + 1) }{1-\delta^2}, \frac{24\delta^2 \beta^2 \|Y^{-1} \|^2 }{1-\delta^2},  0 ] \\
        C_3 & = [24(1-\beta)^2L^2\phi_m(n + 1), 24(1-\beta)^2L^2 \phi_m(n + 1)\frac{\delta^2\alpha^2}{1-\delta^2},\\
        & (1- \beta )^2, 0 ] \\
        C_4 & = [\frac{2\alpha L^2 \phi_m (n+1)}{n}, 0, \frac{\alpha}{n}, 1-\frac{\alpha \mu}{2}],
    \end{aligned}
\end{equation}
and 
\begin{align*}
    \mathbf{r}_k = \begin{bmatrix}
     0 \\
     \frac{8\delta^2}{1-\delta^2} \|Y^{-1} \|^2 (9\alpha^2 L^2 \mathbb{E}[\|\bar{\mathbf{v}}_{k} \|^2] + 3\beta^2 \bar{\nu}^2 n ) \\
     6(1-\beta)^2L^2 \alpha^2  \mathbb{E}[\|\bar{\mathbf{v}}_{k} \|^2] + 2\beta^2\bar{\nu}^2n \\
     0 
    \end{bmatrix}.
\end{align*}
Moreover, let us for convenience denote
\begin{equation}
    \begin{aligned}
        C = \begin{bmatrix}
        C_1^T, \; C_2^T, \; C_3^T, \; C_4^T
        \end{bmatrix}^T.
    \end{aligned}
\end{equation}
It is straighforward to show that
\begin{align*}
    \mathbf{u}_{k+1} \leq C \mathbf{u}_k + \mathbf{r}_k.
\end{align*}
Therefore, if one could find $\x >0$ such that $C \x < \x$, then $\rho (C) <1$. The following lemma provides such a guarantee.


\begin{lemma}\label{lemma6} For the range of $\alpha$ in Lemma~\ref{lemma5}, one can find $\x >0$ such that $\rho (C) \leq 1 - \frac{\alpha \mu}{4}$.
\end{lemma}

It then follows that for $\forall t \in [1, T]$, 
\begin{equation}\label{eq:linear_system}
\begin{aligned}
    \mathbf{u}_t \leq C^t \mathbf{u}_0 + \sum_{k = 0}^{t-1}C^{t-1-k} \mathbf{r}_k. 
\end{aligned}
\end{equation}
In the final stage of the argument, we establish an upper bound on $\mathbf{r}_k$ by bounding $E[\|\bar{\mathbf{v}}_{k} \|^2]$, and show that the bound is independent of the iteration index. 

\begin{lemma}\label{lemma7} Suppose Assumptions \ref{assumption1} - \ref{assumption4} hold, and let $\alpha$ and $\beta$ satisfy the conditions of Theorem \ref{thm:main2}. Then
$E[\|\bar{\mathbf{v}}_{t} \|^2] $ can be upper bounded by a function of $\bar{\nu}$ which is independent of the iteration index $t$.
\end{lemma}

The proof of Theorem \ref{thm:main2} is completed by incorporating the above two lemmas into Lemma~\ref{lemma5} and summarizing the required conditions for the step sizes.
\vspace{-2mm}
\section{Experimental Results}
In this section we report performance of the proposed algorithm, Push-ASGD, in a variety of experimental settings, including three different ML tasks and a numerical study illustrating performance under PL condition. Specifically, Push-ASGD is benchmarked against the following methods tailored to directed networks: SGP/Push-SGD \cite{nedic2016stochastic,assran2019stochastic}, the Push-SAGA algorithm \cite{qureshi2021push} and the Di-CS-SVRG algorithm \cite{chen2021communication}. The first of these algorithm uses only local stochastic gradient updates while the latter two incorporate both global gradient tracking and variance reduction. Please note that there are no theoretical guarantees for the latter two algorithms in decentralized non-convex settings; moreover, those two schemes assume access to IFO while our Push-ASGD is tailored to the more challenging SFO scenario. For all algorithms, the step sizes are selected from $[10^{-7}, 10^{-1}]$ while for Push-ASGD $\beta$ is selected from $[10^{-4}, 10^{-1}]$; the best performance for each method is reported.

Regarding the experiments: we first perform a test on the logistic regression model with non-convex regularization; the second experiment is an image classification task on the CIFAR-10 dataset via a shallow neural network; the third experiment is an NLP classification task on the IMDB dataset via fine-tuning the BERT architecture. We further consider an objective function satisfying the PL condition and numerically test the algorithms' performance.

In all tasks, the training data is distributed randomly without shuffling; hence, nodes are not guaranteed to receive data from all classes. In other words, the nodes experience different data distributions due to a high degree of heterogeneity.

\begin{figure}[H]
\centering
\subfigure[Norm of the gradient (training performance).]{%
  \includegraphics[clip,width=0.5\columnwidth]{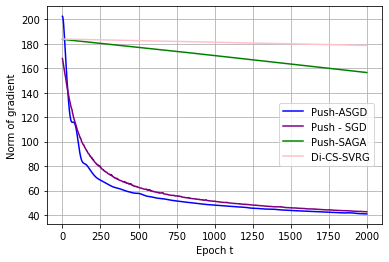}
}

\subfigure[The correct rate (test performance).]{%
  \includegraphics[clip,width=0.5\columnwidth]{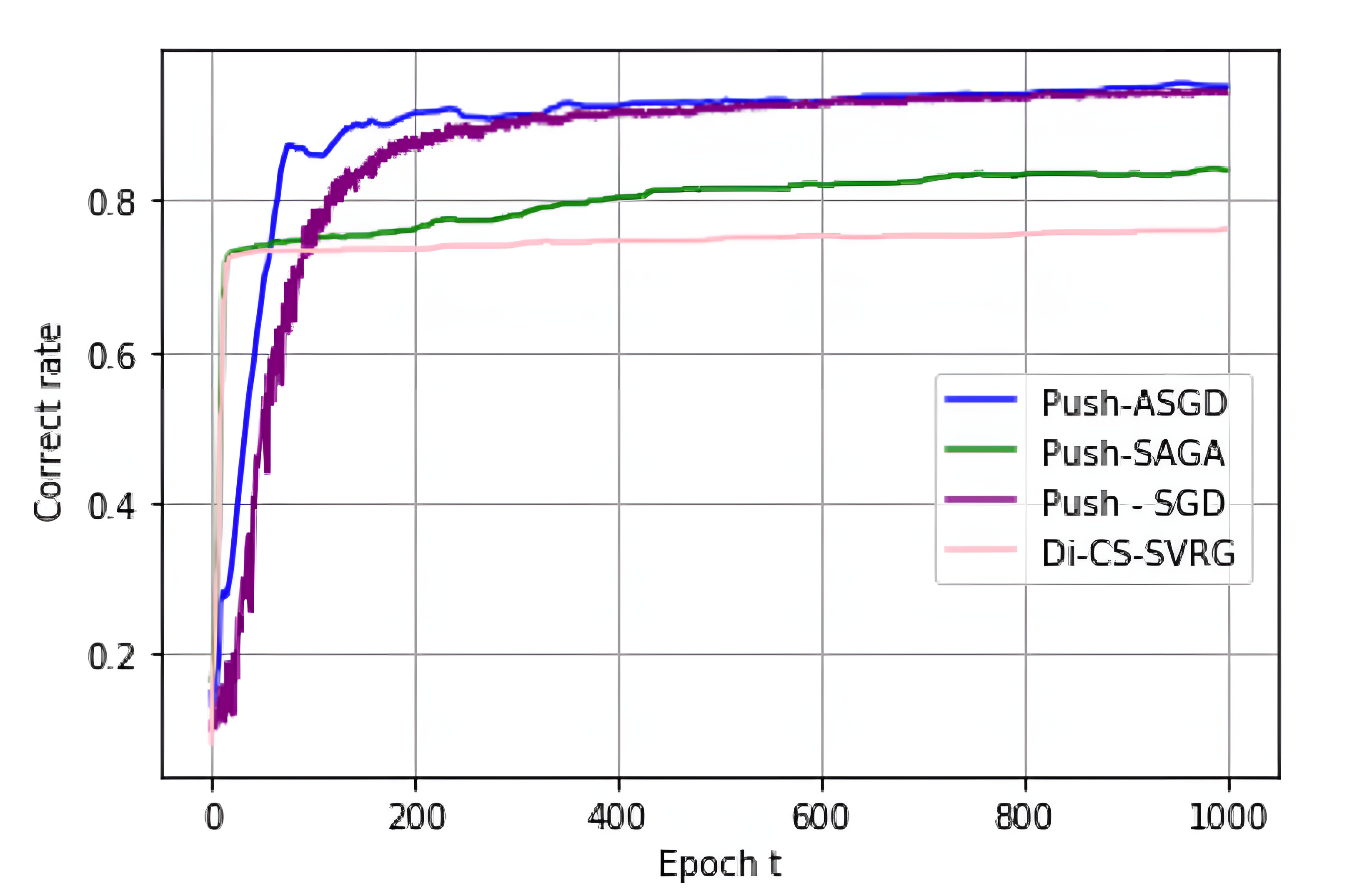}
}
\subfigure[Performance as the network size varies.]{%
  \includegraphics[clip,width=0.5\columnwidth]{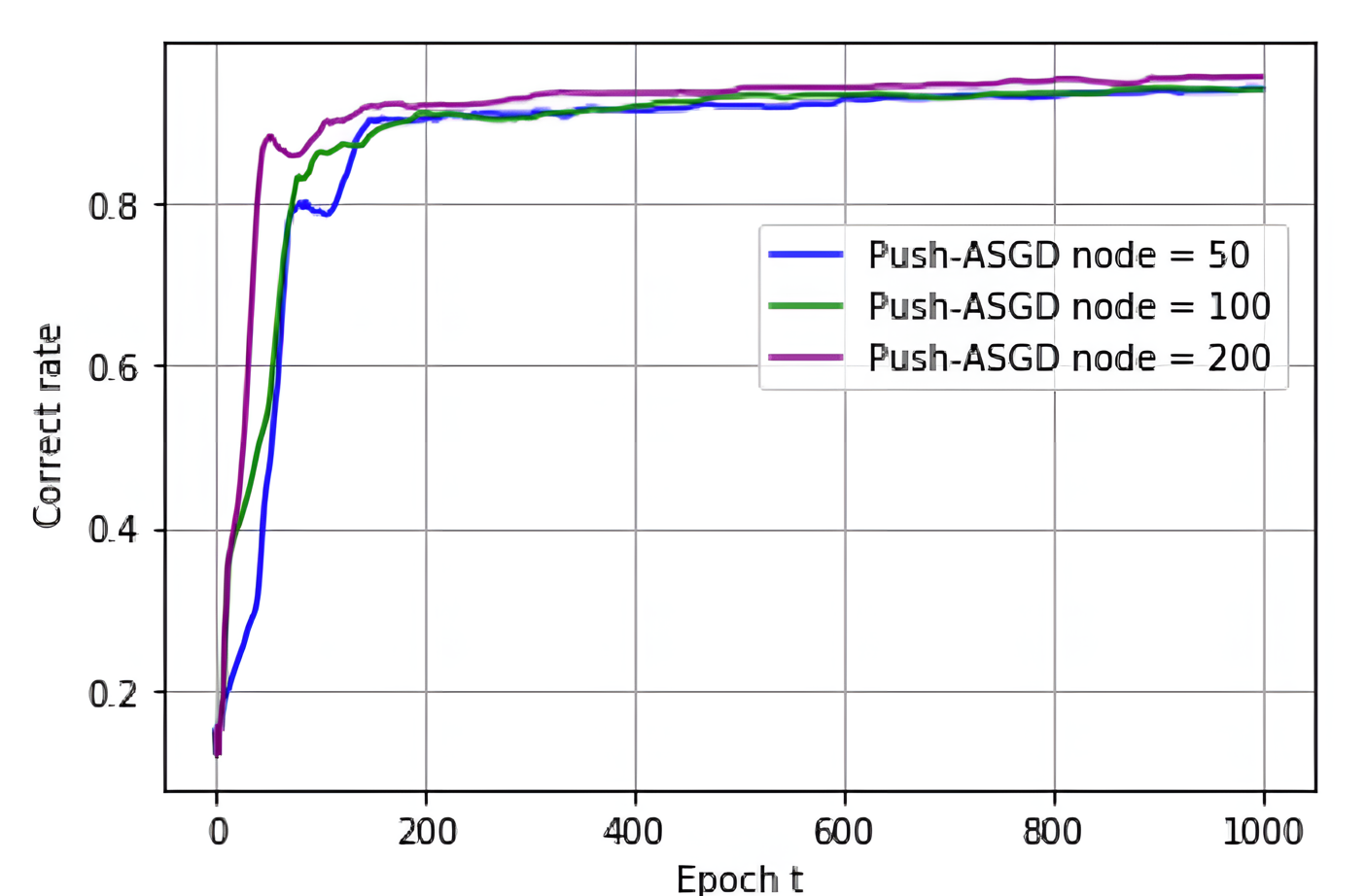}
}
\caption{Performance on MNIST. Push-ASGD achieves lower loss and higher correct rate than the competing schemes.}
\label{fig:logi}
\vspace{-4mm}
\end{figure}

\begin{figure}[H]
\centering
\subfigure[The test loss.]{%
  \includegraphics[clip,width=0.5\columnwidth]{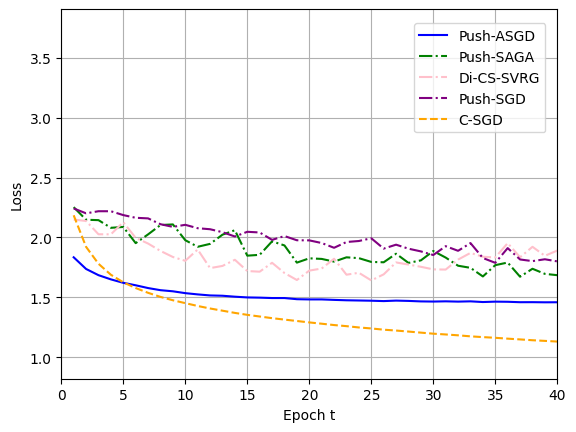}
}

\subfigure[The correct rate.]{%
  \includegraphics[clip,width=0.5\columnwidth]{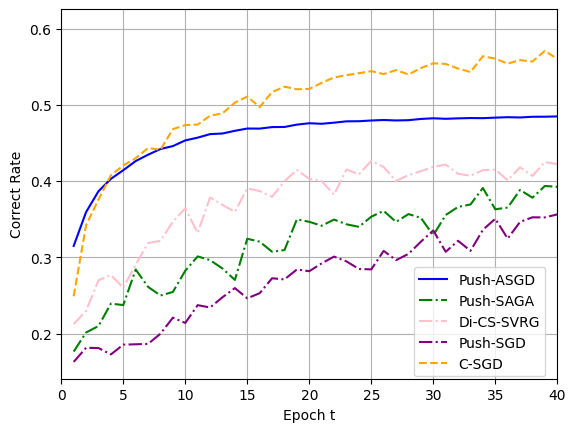}
}
\caption{Performance on CIFAR-10. Push-ASGD achieves lower loss and higher correct rate than the competing schemes.}
\label{fig:cifar}
\vspace{-4mm}
\end{figure}

\begin{figure}[H]
\centering
\subfigure[The test loss.]{%
  \includegraphics[clip,width=0.5\columnwidth]{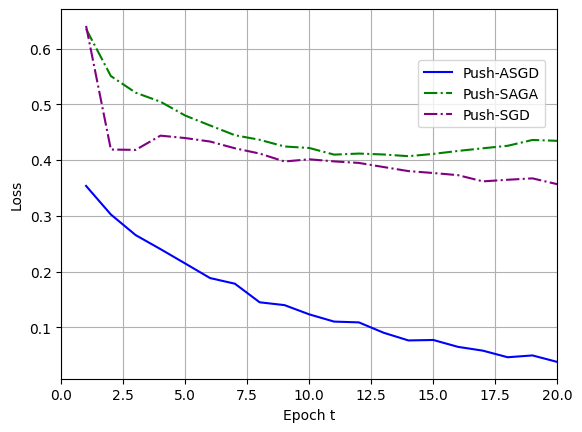}
}

\subfigure[The correct rate.]{%
  \includegraphics[clip,width=0.5\columnwidth]{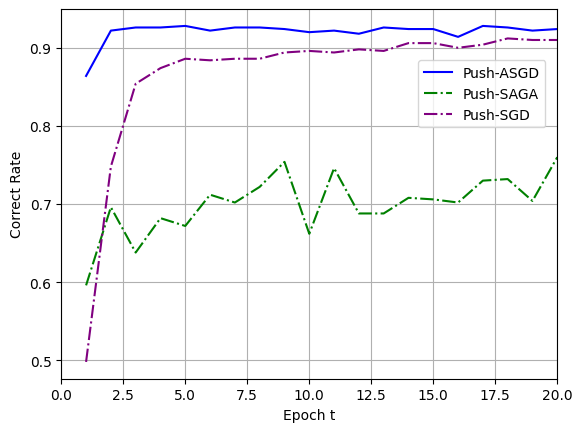}
}
\caption{Performance on the natural language processing task. Push-ASGD achieves lower loss and higher accuracy than the competing schemes.}
\label{fig:bert}
\vspace{-4mm}
\end{figure}

\vspace{-2mm}
\subsection{Non-convex logistic regression}
We first benchmark the performance of Push-ASGD on a decentralized non-convex logistic regression model applied to handwritten digit classification task on the MNIST dataset \cite{lecun1989backpropagation}. The learning task is distributed across $100$ nodes of a time-varying network generated according to the Erd\H{o}s-R\'enyi model and as a directed ring. For the Erd\H{o}s-R\'enyi model, we generate the graph and randomly remove a subset of edges to make the graph directed. The network is switching between the Erd\H{o}s-R\'enyi model, a directed ring and a reversed directed ring. The dataset is distributed such that each node has $12$ images for local training. We deploy a non-convex regularizer and consider the minimization \cite{antoniadis2011penalized}
\begin{equation}
\min_{\x} \left\{\sum_{i=1}^n \sum_{j=1}^N \mathrm{ln}(1+e^{-(\mathbf{m}_{ij}^T\mathbf{x})\y_{ij}}) + \sum_{j=1}^d \frac{\lambda [\mathbf{x}]_j^2}{1 + [\mathbf{x}]_j^2}\right\},
\end{equation}
where $(\mathbf{m}_{ij}, \y_{ij}) $ represents the image feature vector and the corresponding label of the $j$-th image at node $i$. Parameters of the algorithms are set to $\alpha = 6\times 10^{-5}$ and $\beta = 0.015$ (Push-ASGD), $\alpha = 6\times 10^{-5}$ (Subgradient-Push with SGD), and $\alpha =2 \times 10^{-5}$ and $2 \times 10^{-7}$ (Di-CS-SVRG and Push-SAGA, respectively). The regularization parameter is set to $\lambda = 10^{-4}$. For all the experiments in this section, the batch size is set to $1$.

Results of the benchmarking experiments on the non-convex logistic regression task are shown in Figure~\ref{fig:logi}. As can be seen there, the accuracy achieved by Push-ASGD is the highest among all the considered schemes. This confirms our expectation that Push-SGD should outperform Push-SAGA and Di-CS-SVRG since the accuracy of the latter two schemes is adversely affected by the sparse connectivity of the considered directed ring graph structure (such a structure causes instability of primal-dual schemes \cite{towfic2015stability,yuan2020can}).\footnote{A similar phenomenon is observed in Fig.~\ref{fig:bert}; in Fig.~\ref{fig:cifar}, Push-SAGA and Di-CS-SVRG outperform Push-SGD due to a high number of local training data points and reduced stochasticity.}

We also test the performance of Push-ASGD as the network size varies; in particular, we increase the number of network nodes from $50$ to $200$. For fixed network connectivity level and number of local data points, the convergence is faster when there are more nodes in the network (see Figure \ref{fig:logi}(c)).

\vspace{-2mm}
\subsection{Image classification experiments}
Next, we test the performance of the proposed algorithm on a decentralized image classification task involving CIFAR-10 dataset \cite{krizhevsky2012imagenet}. To this end, we rely on a convolutional neural network architecture Lenet \cite{lecun1989backpropagation}. Lenet consists of $5$ layers: two sets of convolutional, activation, and max-pooling layers, followed by two fully-connected layers with activation and a softmax classifier. For this task we utilize the cross-entropy loss.

The time-varying directed network is constructed based on the Erd\H{o}s-R\'enyi model and directed rings. First, the Erd\H{o}s-R\'enyi graph with $10$ nodes is generated, and then several edges are removed to induce a directed graph. Each node is assigned $5000$ images from the CIFAR-10 dataset for local training. For all algorithms, step size is set to $\alpha = 10^{-2}$; for Push-ASGD, the momentum step size is set to $\beta = 0.05$. As a reference, we also provide a comparison with the centralized stochastic gradient descent (C-SGD), for which the training data includes all $50000$ images. For all the experiments in this section, the batch size is $32$.

The test loss and the correct rate are reported in Figures \ref{fig:cifar}(a) and  \ref{fig:cifar}(b), respectively. As seen there, the proposed algorithm, Push-ASGD, outperforms other decentralized schemes. The gap between Push-ASGD and C-SGD is due to the impact of distributing the dataset across the network nodes while maintaining the same total amount of data as used by the centralized method.
\vspace{-2mm}
\subsection{Natural language processing experiments}
The remaining real-world data experiment involves an NLP classification task via fine-tuning a deep learning language model. In particular, we train this model on the IMDB dataset that contains the texts of reviews and the corresponding binary tags implying whether the review is positive or negative \cite{maas2011learning}. We still consider an Erd\H{o}s-R\'enyi-based directed time-varying network and distribute the IMDB training data such that each node has access to $2000$ reviews and uses them for local training. The model is constructed by adding a linear classification layer to the pre-trained Bidirectional Encoder Representations from Transformers (BERT) \cite{devlin2018bert} architecture, and fine-tuned locally. We again utilize the cross-entropy loss.

The Push-ASGD and Push-SGD algorithms use step size $\alpha = 0.002$, while Push-SAGA uses a smaller step size $\alpha = 0.0002$ to avoid divergence. For the Push-ASGD algorithm, the momentum step size is set to $\beta = 0.025$.  In all the experiments in this section the batch size is set to $4$. The performance of the algorithms is shown in Fig.~\ref{fig:bert}. As seen there, Push-ASGD achieves lower loss and converges to the highest correct rate of over $90\%$.\footnote{The performance of Di-CS-SVRG is omitted since it failed to converge in 20 epochs.} 
\vspace{-2mm}
\subsection{The PL condition}
\begin{figure}[H]
\centering
  \includegraphics[clip,width=0.5\columnwidth]{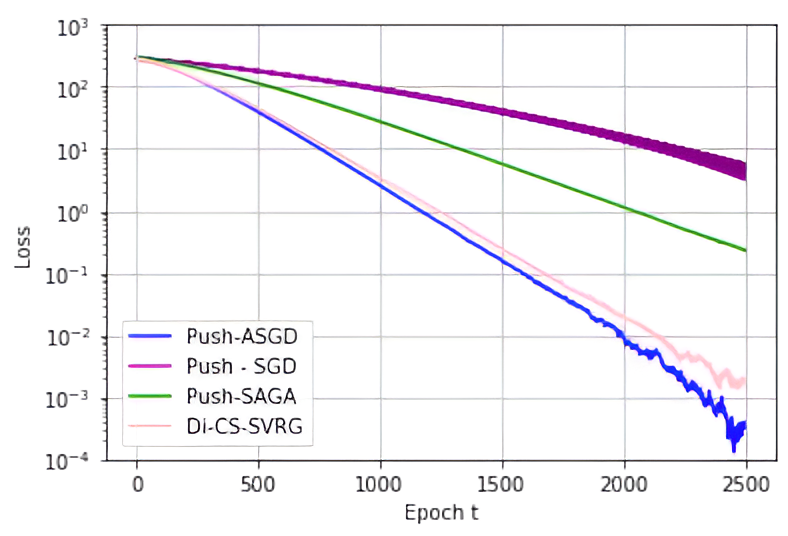}
\caption{In simulations of a setting where PL condition holds, Push-ASGD converges faster than other benchmarking algorithms.}
\label{fig:PL}
\vspace{-4mm}
\end{figure}

Lastly, we test the performance of the algorithms in an application to minimizing a global function that satisfies the PL condition. The local functions is defined as $f_i(x) = x^2 + 3\sin^2(x) + a_i \cos(x)$, where $a_i$ are non-zero parameters satisfying $\sum_{i=1}^n a_i = 0$ so that the global function is $ F(x) = x^2 + 3\sin^2(x)$. The global function is non-convex and satisfies the PL condition \cite{karimi2016linear}. To simulate stochastic gradient, we add random Gaussian noise with mean $0$ and standard deviation $1/2$ to the gradient at each node. The network consists of $100$ nodes; similar to the previous experiments, the time-varying directed network topology is based on the Erd\H{o}s-R\'enyi model and directed rings. 
As seen in Fig. \ref{fig:PL}, Push-ASGD converges the fastest, followed by the two benchmarking algorithms with gradient acceleration; Push-SGD, computing a simple stochastic gradient, converges the slowest and is less stable than other algorithms.

\vspace{-2mm}
\section{Conclusion}
The paper presents the first analytical study of decentralized stochastic non-convex optimization over time-varying directed graphs, and introduces
a novel stochastic optimization algorithm for this problem. The method, Push-ASGD, is the first scheme that achieves the 
SFO complexity of $\O(1/\epsilon^{1.5})$ for smooth objectives; in addition, it enjoys linear convergence under the PL condition. Push-ASGD relies on a push-sum protocol to perform local aggregation under communication asymmetry, while employing a novel stochastic gradient estimator to deal with uncertainties stemming from noise and heterogeneity in local data. The proposed gradient estimator incorporates momentum-based variance reduction and gradient tracking techniques to recursively estimate global gradient, which is unknown to the participating agents. Extensive experiments demonstrate that Push-ASGD outperforms existing methods for distributed optimization over time-varying directed networks. 



\section*{Appendix}
In the appendix we provide details of the analysis summarized in the theorems.
\vspace{-2mm}
\subsection{Proof of Lemma~1}
We start by defining the norm with respect to time-varying stochastic vectors, $\phi_{t}$. Recall the update of ${\z}_t$, 
\begin{align*}
    \z_{t+1} & = \Tilde{W}^{(t)} (\z_t - \alpha \h_t),
\end{align*}
where $\Tilde{W}^{(t)} = \Tilde{W}_m^{(t)} \otimes I_d $ and $\z_t, \h_t \in \R^{nd} $. The $(i, j)$-th entry in the column-stochastic matrix $ W_m^{(t)}$ is given by
\begin{equation}
    w_{ij}^{(t)} = \frac{1}{d_j^{out, t}+1} \quad \mathrm{for} \quad (j, i) \in \E(t),
\end{equation} 
where $d_j^{out, t}$ denotes the out-degree of agent $j$ at time $t$. Since the entries of $\Tilde{W}_m^{(t)}$ are given by $\Tilde{w}_{i, j}^{(t)} = w_{i, j}^{(t)} y^j_{t}/y^i_{t+1}$, it is guaranteed that each row of $\Tilde{W}_m^{(t)} $ has row sum equal to $1$ \cite{nedic2017achieving}. Since $\Tilde{W}_m^{(t)}$ is row-stochastic and the network is strongly connected, $\Tilde{W}_m^{(t)}$ can also be viewed as a row-stochastic mixing matrix for strongly-connected directed graphs. Using the weight policy of $W_m^t $ and the update of $y_t$, we can derive that the positive entries in $\Tilde{W}_m^{(t)}$ can be uniformly lower bounded as $\Tilde{w}_{i, j}^{(t)} \geq \omega = \frac{1}{n^{n+2}} $ for $(j, i) \in \E(t) $ \cite{nedic2017achieving}.
Consider the stochastic vector sequence $\{\phi_t \}$ such that ${\phi_{t+1}}^T \Tilde{W}_m^{(t)} = \phi_t^T $; from Lemma 3.3 of \cite{nedich2022ab}, the sequence $\{\phi_t \}$ exists and has element-wise lower bound, i.e., $[\phi_t]_i \geq \frac{\omega^n}{n}$. We recall the definition of $\mathfrak{L}^2(\z_t, \phi_t)$, 
\begin{align*}
    \mathfrak{L}^2(\z_t, \phi_t) = \|(diag(\phi_t)^{\frac{1}{2}} \otimes I_d )(\z_t - \hat{\z}_t ) \|_F^2,
\end{align*}
where $\hat{\z}_t$ is the $\phi_t$-weighted average of $\z_t$ in $\mathcal{R}^{nd}$.
Following Lemma 4.2 in \cite{nedich2022ab},
we can obtain $\mathfrak{L}(\Tilde{W}_m^{(t)} \z, \phi_{t+1}) \leq \lambda_t \mathfrak{L} (\z, \phi_t ) $, where $\lambda_t = \sqrt{1 - \frac{\min_i([\phi_{t+1}]_i)\omega^2}{\max_i([\phi_t]_i) (n-1)^2} } \in (0, 1)$.
Using the triangle and Young's inequalities we obtain
\begin{eqnarray}
\begin{split}
    \mathfrak{L}^2(\z_{t+1}, \phi_{t+1}) & = \mathfrak{L}^2(\Tilde{W}^{(t)} (\z_t - \alpha \h_t), \phi_{t+1} ) \\
    & \leq  (1+r) \mathfrak{L}^2(\Tilde{W}^{(t)} \z_t, \phi_{t+1} ) + (1+\frac{1}{r}) \alpha^2 \mathfrak{L}^2(\Tilde{W}^{(t)}  \h_t, \phi_{t+1} ) \\
    & \leq (1+r)\delta^2 \mathfrak{L}^2(\z_t, \phi_t) + (1+\frac{1}{r}) \delta^2 \alpha^2 \mathfrak{L}^2(\h_t, \phi_t) \\
    & \leq  \frac{1+\delta^2}{2}  \mathfrak{L}^2(\z_t, \phi_t) + \frac{(1+\delta^2)\delta^2 \alpha^2}{1-\delta^2}  \mathfrak{L}^2(\h_t, \phi_t) \\
    & \leq  \frac{1+\delta^2}{2}  \mathfrak{L}^2(\z_t, \phi_t) + \frac{2\delta^2 \alpha^2}{1-\delta^2}  \mathfrak{L}^2(\h_t, \phi_t), 
\end{split}
\label{eq_1}
\end{eqnarray}
where $r = \frac{1-\delta^2}{2\delta^2}$ and $\delta = \max_{t \geq 0} \lambda_t $. The proof of Lemma \ref{lemma1} is completed by taking the expectation of both sides of the inequality.

\vspace{-2mm}
\subsection{Proof of Lemma~2}
We start by applying techniques similar to those used in the consensus error analysis in the proof of Lemma \ref{lemma1}, yielding
\begin{align*}
    \mathfrak{L}^2({\h}_{t+1}, \phi_{t+1} ) 
    & = \mathfrak{L}^2 (\Tilde{W}^{(t)} \h_t + \Tilde{W}^{(t)} (Y^{-1}_t \otimes I_d) (\v_{t+1} - \v_t), \phi_{t+1} ) \\
    & \leq \frac{1+\delta^2}{2} \mathfrak{L}^2({\h}_t, \phi_{t}) + \frac{2\delta^2 }{1-\delta^2} \|Y^{-1} \|^2 \mathfrak{L}^2(\v_{t+1} - \v_t, \phi_{t+1} ) \\
    & \leq \frac{1+\delta^2}{2} \mathfrak{L}^2({\h}_t, \phi_t) + \frac{8\delta^2 }{1-\delta^2} \|Y^{-1} \|^2 \|\v_{t+1} - \v_t \|^2_2
\end{align*}
where $\|Y^{-1} \| = \sup_{t} \|Y_t^{-1} \|_{\max}$ denotes the supremum of the inverse elements across all iterations, and the second inequality follows from the definition of the $\mathcal{L}$-norm.
Moreover, $\|Y^{-1} \| \leq n^n$ since the smallest element of $\|Y_t\|_{\max} \geq \frac{1}{n^n} $ for any $t$ \cite{nedic2017achieving}. After taking expectation of both sides, we obtain
\begin{eqnarray}\label{eq_2}
\mathbb{E}\mathfrak{L}^2({\h}_{t+1}, \phi_{t+1} ) 
& \leq & \frac{1+\delta^2}{2} \mathbb{E}\mathfrak{L}^2({\h}_{t}, \phi_{t})
\nonumber \\
& + & \frac{8\delta^2 \|Y^{-1} \|^2}{1-\delta^2} \mathbb{E}\|\v_{t+1} - \v_t \|^2_2.
\end{eqnarray}
We proceed by deriving an upper bound on $\mathbb{E}\|\mathbf{v}_{t+1} - \mathbf{v}_t \|^2_2 $. To this end, note that for each $i$,
\begin{align*}
    \mathbf{v}^i_{t+1} - \mathbf{v}^i_t & = \nabla f_i(\mathbf{z}^i_{t+1}, \mathbf{\xi}^i_{t}) + (1-\beta)(\mathbf{v}^i_{t} -\nabla f_i(\mathbf{z}^i_{t}, \mathbf{\xi}^i_{t} ) ) - \mathbf{v}^i_t \\
    & = \nabla f_i(\mathbf{z}^i_{t+1}, \mathbf{\xi}^i_{t}) - \nabla f_i(\mathbf{z}^i_{t}, \mathbf{\xi}^i_{t} ) - \beta \mathbf{v}^i_t + \beta \nabla f_i(\mathbf{z}^i_{t}, \mathbf{\xi}^i_{t} ) \\
    & = \nabla f_i(\mathbf{z}^i_{t+1}, \mathbf{\xi}^i_{t}) - \nabla f_i(\mathbf{z}^i_{t}, \mathbf{\xi}^i_{t} )\\ & - \beta (\mathbf{v}^i_t - \nabla \mathbf{f}_i(\mathbf{z_t^i}) ) + \beta (\nabla f_i(\mathbf{z}^i_{t}, \mathbf{\xi}^i_{t} ) - \nabla \mathbf{f}_i(\mathbf{z_t^i})  ).
\end{align*}
Taking the expectation of $\|\mathbf{v}^i_{t+1} - \mathbf{v}^i_t \|^2 $, 
\begin{align*}
    \mathbb{E}[\|\mathbf{v}^i_{t+1} - \mathbf{v}^i_t \|^2 ]  & \stackrel{(a)}{ \leq} 3\mathbb{E}[\| \nabla f_i(\mathbf{z}^i_{t+1}, \mathbf{\xi}^i_{t}) - \nabla f_i(\mathbf{z}^i_{t}, \mathbf{\xi}^i_{t} ) \|^2 ] +\\
    & \quad 3\beta^2 \mathbb{E}[\| \mathbf{v}^i_t - \nabla \mathbf{f}_i(\mathbf{z_t^i}) \|^2 ] + 3\beta^2 \mathbb{E}[\| \nabla f_i(\mathbf{z}^i_{t}, \mathbf{\xi}^i_{t} ) - \nabla \mathbf{f}_i(\mathbf{z_t^i}) \|^2 ] \\
    & \stackrel{(b)}{ \leq} 3L^2\mathbb{E}[\|\mathbf{z}^i_{t+1} - \mathbf{z}^i_t \|^2] + 3\beta^2 \mathbb{E}[\| \mathbf{v}^i_t - \nabla \mathbf{f}_i(\mathbf{z_t^i}) \|^2 ] + 3\beta^2 \nu_i^2,
\end{align*}
where $(a)$ is due to Cauchy–Schwarz inequality while for $(b)$ we invoke the smoothness assumption; here $L$ denotes the smoothness parameter and $\nu_i^2$ is a bound on the stochastic gradient estimate for given $i$. 
Summing up from $i = 1$ to $n$ yields
\begin{align*}
    \mathbb{E}[\|\mathbf{v}_{t+1} - \mathbf{v}_t \|^2 ] & \stackrel{(c)}{ \leq}  3L^2\mathbb{E}[\|\mathbf{z}_{t+1} - \mathbf{z}_t \|^2] + 3\beta^2 \mathbb{E}[\| \mathbf{v}_t - \nabla \mathbf{f}(\mathbf{z_t}) \|^2 ] + 3\beta^2 \bar{\nu}^2 n \\
    & = 3L^2\mathbb{E}[\|\mathbf{z}_{t+1} - \bar{\mathbf{x}}_{t+1} +\bar{\mathbf{x}}_{t+1} - \bar{\mathbf{x}}_{t} + \bar{\mathbf{x}}_{t}  - \mathbf{z}_t \|^2] \\ & + 3\beta^2 \mathbb{E}[\| \mathbf{v}_t - \nabla \mathbf{f}(\mathbf{z_t}) \|^2 ]  + 3\beta^2 \bar{\nu}^2 n \\
    & \stackrel{(d)}{ \leq}  3L^2[3\mathbb{E}[\|\bar{\mathbf{x}}_{t+1} - \bar{\mathbf{x}}_{t}  \|^2] + 3\mathbb{E}[\|\mathbf{z}_{t+1} - \bar{\mathbf{x}}_{t+1} \|^2 + \|\mathbf{z}_{t} - \bar{\mathbf{x}}_{t} \|^2 ]]  \\
    &  + 3\beta^2 \mathbb{E}[\| \mathbf{v}_t - \nabla \mathbf{f}(\mathbf{z_t}) \|^2 ] + 3\beta^2 \bar{\nu}^2 n \\
    & \stackrel{(e)}{ \leq}  3L^2[3\alpha^2 \mathbb{E}[\|\bar{\mathbf{v}}_{t} \|^2] + 3\mathbb{E}[\|\mathbf{z}_{t+1} - \bar{\mathbf{x}}_{t+1} \|^2 + \|\mathbf{z}_{t} - \bar{\mathbf{x}}_{t} \|^2 ]]  \\
    & + 3\beta^2 \mathbb{E}[\| \mathbf{v}_t - \nabla \mathbf{f}(\mathbf{z_t}) \|^2 ] + 3\beta^2 \bar{\nu}^2 n,
\end{align*}
where $\bar{\mathbf{v}}_t = [(\frac{1}{n}\sum_j \v_t^j)^T, \cdots, (\frac{1}{n}\sum_j \v_t^j)^T]^T \in \R^{np}$ and $\bar{\nu}^2 =\frac{1}{n} \sum_{i=1}^n \nu_i^2$; note that $(c)$ follows $(b)$, $(d)$ is due to the Cauchy-Schwarz inequality and $(e)$ stems from the update rule of $\x_t$. The term $\mathbb{E}\|\mathbf{z}_{t+1} - \bar{\mathbf{x}}_{t+1} \|^2$ appears in the upper bound and thus deserves closer attention.
Recall that $\hat{\z}_t$ is the $\phi_t$-weighted average of $\z_t$; letting $\phi_m = d/\min_{t, i} [\phi_t]_i$, it holds that
\begin{align}
    & \|\mathbf{z}_t - \bar{\mathbf{x}}_t \|^2  = \|\mathbf{z}_t - \hat{\z}_t + \hat{\z}_t - \bar{\mathbf{x}}_t \|^2_F \nonumber \\ & \stackrel{(f)}{ \leq} 2 \frac{1}{\min(\phi_t)} \mathfrak{L}^2({\z}_t, \phi_t ) + 2n \| \sum_{j=1}^n [\phi_t]_j\z_t^j - \frac{1}{n}\sum_{j=1}^n [y_t]_j \z_t^j \|_2^2 \nonumber \\
    & = 2 \frac{1}{\min(\phi_t)} \mathfrak{L}^2({\z}_t, \phi_t ) + 2n\sum_{m=1}^d  [\sum_{j=1}^n [\phi_t]_j\z_t^j - \frac{1}{n}\sum_{j=1}^n [y_t]_j \z_t^j ]_m^2 \nonumber \\
    & \stackrel{(g)}{ \leq} 2 \frac{1}{\min(\phi_t)} \mathfrak{L}^2({\z}_t, \phi_t ) + 2n\sum_{m=1}^d \max_i [\sum_{j=1}^n [\phi_t]_j\z_t^j -  \z_t^i ]_m^2 \nonumber \\
    & \stackrel{(h)}{ \leq} 2 \frac{1}{\min(\phi_t)} \mathfrak{L}^2({\z}_t, \phi_t ) + 2n\sum_{m=1}^d \max_i \| \sum_{j=1}^n [\phi_t]_j \z_t^j - \z_t^i \|_2^2 \nonumber \\
    & =  2\frac{1}{\min(\phi_t)} \mathfrak{L}^2({\z}_t, \phi_t ) + 2nd \max_i \| \sum_{j=1}^n [\phi_t]_j \z_t^j - \z_t^i \|_2^2 \nonumber \\
    & \stackrel{(i)}{ \leq} 2\frac{1}{\min(\phi_t)} \mathfrak{L}^2({\z}_t, \phi_t ) + 2nd \| \z_t - \hat{\z}_t \|_2^2 \nonumber \\
    & \stackrel{(j)}{ \leq} 2\phi_m\mathfrak{L}^2({\z}_t, \phi_t ) + 2n\phi_m \mathfrak{L}^2({\z}_t, \phi_t ), \label{z-x}
\end{align}
\noindent where
$(f)$ is due to the Cauchy-Schwarz inequality and the fact that column-stochasticity of matrix $W_m^{(t)}$ ensures that $\sum_{j=1}^n[y_t]_j = \sum_{j=1}^n[y_0]_j = n$.  
One can show (g) by first finding the maximum and minimum of the weighted sum $\sum_{i=1}^n a_i [\z_t^i]_m$ subject to $\sum_{i=1}^n a_i = 1, a_i \geq 0$. Specifically, the maximum is achieved by putting all the weight on the largest $[\z_t^i]_m$, while the minimum is achieved by putting all the weight on the smallest $[\z_t^i]_m$. Let $i^* = \mathrm{argmax}_i [\z_t^i]_m$ and $j^* = \mathrm{argmin}_i [\z_t^i]_m$; then $\max_{\{a_i\}} [\sum_{j=1}^n [\phi_t]_j\z_t^j - \sum_{j=1}^n a_j \z_t^j ]_m^2 = \max \{[\sum_{j=1}^n [\phi_t]_j\z_t^j - \z_t^{i^*} ]_m^2, [\sum_{j=1}^n [\phi_t]_j\z_t^j - \z_t^{j^*} ]_m^2 \}  $. Therefore, it must be that $ [\sum_{j=1}^n [\phi_t]_j\z_t^j - \sum_{j=1}^n a_j \z_t^j ]_m^2 \leq \max_i  [\sum_{j=1}^n [\phi_t]_j\z_t^j - \z_t^i ]_m^2 $.
$(h)$ is due to the fact that the square of each entry of a vector is no greater than the squared $\ell_2$ norm of the vector. Finally, $(i)-(j)$ are due to the definition of $\mathfrak{L}^2({\z}_t, \phi_t ) $. Taking the expectation and revisiting the bound on $\mathbb{E}\|\mathbf{v}_{t+1} - \mathbf{v}_t \|^2_2 $ yields
\begin{align*}
    \mathbb{E}[\|\mathbf{v}_{t+1} - \mathbf{v}_t \|^2 ] 
    & = 3L^2[3\alpha^2 \mathbb{E}[\|\bar{\mathbf{v}}_{t} \|^2] + 3\mathbb{E}[\|\mathbf{z}_{t+1} - \bar{\mathbf{x}}_{t+1} \|^2 + \|\mathbf{z}_{t} - \bar{\mathbf{x}}_{t} \|^2 ]]  \\
     & + 3\beta^2 \mathbb{E}[\| \mathbf{v}_t - \nabla \mathbf{f}(\mathbf{z_t}) \|^2 ] + 3\beta^2 \bar{\nu}^2 n \\
     & \leq 3L^2[3\alpha^2 \mathbb{E}[\|\bar{\mathbf{v}}_{t} \|^2] \\
     & + 6\phi_m(n+1)\mathbb{E}[ \mathfrak{L}^2({\z}_{t+1}, \phi_{t+1} ) + \mathfrak{L}^2({\z}_t, \phi_{t} )]]  \\
     & + 3\beta^2 \mathbb{E}[\| \mathbf{v}_t - \nabla \mathbf{f}(\mathbf{z_t}) \|^2 ] + 3\beta^2 \bar{\nu}^2 n,
\end{align*}
where the last inequality is due to $(e)$ and $(i)$.
Substituting the bound on $\mathbb{E}[\|\mathbf{v}_{t+1} - \mathbf{v}_t \|^2 ]$ in (\ref{eq_2}), we obtain
\begin{align*}
    \mathbb{E} [\mathfrak{L}^2({\h}_{t+1}, \phi_{t+1} )] 
    & \leq \frac{1+\delta^2}{2} \mathbb{E}[ \mathfrak{L}^2({\h}_{t}, \phi_{t} )] + \frac{8\delta^2 }{1-\delta^2} \|Y^{-1} \|^2 \mathbb{E} \|\v_{t+1} - \v_t \|^2 ]  \\
    & \leq \frac{1+\delta^2}{2} \mathbb{E}[ \mathfrak{L}^2({\h}_{t}, \phi_t )] + \frac{8\delta^2 }{1-\delta^2} \|Y^{-1} \|^2 [ 3L^2[3\alpha^2 \mathbb{E}[\|\bar{\mathbf{v}}_{t} \|^2] \\ & + 6\phi_m(n+1)\mathbb{E}[ \mathfrak{L}^2({\z}_{t+1}, \phi_{t+1} ) + \mathfrak{L}^2({\z}_t, \phi_t )]] \\
    & + 3\beta^2 \mathbb{E}[\| \mathbf{v}_t - \nabla \mathbf{f}(\mathbf{z_t}) \|^2 ] + 3\beta^2 \bar{\nu}^2 n  ].
\end{align*}
\vspace{-2mm}
\subsection{Proof of Lemma~3}
Recall the rule for updating the local gradient estimate $\mathbf{v}_t^i$,
\begin{eqnarray*}
\mathbf{v}^i_{t} &=& \nabla f_i(\mathbf{z}^i_{t}, \mathbf{\xi}^i_{t-1}) + (1-\beta)(\mathbf{v}^i_{t-1} -\nabla f_i(\mathbf{z}^i_{t-1}, \mathbf{\xi}^i_{t-1} ) ) \\
& = & \beta \nabla f_i(\mathbf{z}^i_{t}, \mathbf{\xi}^i_{t-1}) \\
& + & (1-\beta)(\mathbf{v}^i_{t-1}
+ \nabla f_i(\mathbf{z}^i_{t}, \mathbf{\xi}^i_{t-1})  -\nabla f_i(\mathbf{z}^i_{t-1}, \mathbf{\xi}^i_{t-1} )). 
\end{eqnarray*}
For all $t \geq 1$ and $i$,

\begin{align*}
    & \mathbf{v}^i_{t} - \nabla \mathbf{f}_i(\mathbf{z^i_t})  = \beta \nabla f_i(\mathbf{z}^i_{t}, \mathbf{\xi}^i_{t-1})  + (1-\beta)(\mathbf{v}^i_{t-1} + \nabla f_i(\mathbf{z}^i_{t}, \mathbf{\xi}^i_{t-1})  \\
    & -\nabla f_i(\mathbf{z}^i_{t-1}, \mathbf{\xi}^i_{t-1} )  ) - \beta \nabla \mathbf{f}_i(\mathbf{z^i_t}) - (1-\beta) \nabla \mathbf{f}_i(\mathbf{z^i_t}) \\
    & = \beta (\nabla f_i(\mathbf{z}^i_{t}, \mathbf{\xi}^i_{t-1}) - \nabla \mathbf{f}_i(\mathbf{z^i_t}) ) + (1-\beta) (\mathbf{v}^i_{t-1} + \nabla f_i(\mathbf{z}^i_{t}, \mathbf{\xi}^i_{t-1}) \\
    & -\nabla f_i(\mathbf{z}^i_{t-1}, \mathbf{\xi}^i_{t-1} ) -\nabla \mathbf{f}_i(\mathbf{z^i_t}) )  \\
    & = \beta (\nabla f_i(\mathbf{z}^i_{t}, \mathbf{\xi}^i_{t-1}) - \nabla \mathbf{f}_i(\mathbf{z^i_t}) ) \\
    & + (1-\beta) (\nabla f_i(\mathbf{z}^i_{t}, \mathbf{\xi}^i_{t-1})  -\nabla f_i(\mathbf{z}^i_{t-1}, \mathbf{\xi}^i_{t-1} ) \\
    & + \nabla \mathbf{f}_i(\mathbf{z^i_{t-1}}) -\nabla \mathbf{f}_i(\mathbf{z^i_t}) )  + (1-\beta) (\mathbf{v}^i_{t-1} - \nabla \mathbf{f}_i(\mathbf{z^i_{t-1}})). 
\end{align*}

Let $\mathbf{s}_t =\sum_i \nabla f_i(\mathbf{z}^i_{t}, \mathbf{\xi}^i_{t-1}) - \nabla \mathbf{f}_i(\mathbf{z^i_t})$ and \\
$\mathbf{s}'_t = \sum_i \nabla f_i(\mathbf{z}^i_{t}, \mathbf{\xi}^i_{t-1})  -\nabla f_i(\mathbf{z}^i_{t-1}, \mathbf{\xi}^i_{t-1} ) + \nabla \mathbf{f}_i(\mathbf{z^i_{t-1}}) -\nabla \mathbf{f}_i(\mathbf{z^i_t})  $.

Summing over $i$ from $1$ to $n$ and taking the expectation conditioned on $\mathcal{F}_t$ gives
\begin{align*}
    \mathbb{E}[\| \mathbf{v}_t - \nabla \mathbf{f}(\mathbf{z_t}) \|^2 |\mathcal{F}_t] & = (1 - \beta)^2 \|\mathbf{v}_{t-1} - \nabla \mathbf{f}(\mathbf{z_{t-1}}) \|^2 + \mathbb{E}[\|\beta \mathbf{s}_t + (1-\beta)\mathbf{s}'_t \|^2 |\mathcal{F}_t] \\
    & + 2\mathbb{E}[ \langle (1-\beta)(\mathbf{v}_{t-1} - \nabla \mathbf{f}(\mathbf{z_{t-1}}) ), \beta \mathbf{s}_t + (1-\beta)\mathbf{s}'_t \rangle | \mathcal{F}_t] \\
    & \stackrel{(3a)}{ \leq} (1 - \beta)^2 \|\mathbf{v}_{t-1} - \nabla \mathbf{f}(\mathbf{z_{t-1}}) \|^2 + \mathbb{E}[\|\beta \mathbf{s}_t + (1-\beta)\mathbf{s}'_t \|^2 |\mathcal{F}_t] \\
    & \stackrel{(3b)}{ \leq} (1 - \beta)^2 \|\mathbf{v}_{t-1} - \nabla \mathbf{f}(\mathbf{z_{t-1}}) \|^2 + 2\beta^2 \mathbb{E}[\|\mathbf{s}_t \|^2 | \mathcal{F}_t] \\
    & + 2(1-\beta)^2 \mathbb{E}[\|\mathbf{s}'_t \|^2 | \mathcal{F}_t] \\
    & \stackrel{(3c)}{ =} (1 - \beta)^2 \|\mathbf{v}_{t-1} - \nabla \mathbf{f}(\mathbf{z_{t-1}}) \|^2 + 2\beta^2 \mathbb{E}[\|\mathbf{s}_t \|^2 | \mathcal{F}_t] \\
    & + 2(1-\beta)^2 \mathbb{E}[\|\sum_i \nabla f_i(\mathbf{z}^i_{t}, \mathbf{\xi}^i_{t-1})  -\nabla f_i(\mathbf{z}^i_{t-1}, \mathbf{\xi}^i_{t-1} ) \|^2 | \mathcal{F}_t],
\end{align*}
where $(3a)$ is due to Assumption \ref{assumption1}, $(3b)$ is due to the Cauchy-Schwarz inequality and $(3c)$ is due to the conditional variance decomposition. The upper bound on the unconditional expectation can then be derived as

\begin{equation}
\begin{aligned}
    \mathbb{E}[\| \mathbf{v}_t - \nabla \mathbf{f}(\mathbf{z_t}) \|^2 ] & \leq (1 - \beta)^2 \mathbb{E}[\|\mathbf{v}_{t-1} - \nabla \mathbf{f}(\mathbf{z_{t-1}}) \|^2] + 2\beta^2 n \bar{\nu}^2 \\
    & + 2(1-\beta)^2L^2 \mathbb{E}[\|\mathbf{z}_t - \mathbf{z}_{t-1} \|^2 ] \\
    & \stackrel{(3d)}{ \leq} (1 - \beta)^2 \mathbb{E}[ \|\mathbf{v}_{t-1} - \nabla \mathbf{f}(\mathbf{z_{t-1}}) \|^2] + 2\beta^2 n \bar{\nu}^2  + 6(1-\beta)^2 \\ &  L^2  (\mathbb{E}[\|\mathbf{z}_t - \bar{\mathbf{x}}_t \|^2 + \|\bar{\mathbf{x}}_t-\bar{\mathbf{x}}_{t-1} \|^2 + \|\mathbf{z}_{t-1} - \bar{\mathbf{x}}_{t-1} \|^2 ] ) \\
    & = (1 - \beta)^2 \mathbb{E}[\|\mathbf{v}_{t-1} - \nabla \mathbf{f}(\mathbf{z_{t-1}}) \|^2] + 2\beta^2 n \bar{\nu}^2 \\
    & + 6(1-\beta)^2L^2 \alpha^2 \mathbb{E}[\|\bar{\mathbf{v}}_{t-1} \|^2] \\
    & + 6(1-\beta)^2L^2 (\mathbb{E}[\|\mathbf{z}_t - \bar{\mathbf{x}}_t \|^2 + \|\mathbf{z}_{t-1} - \bar{\mathbf{x}}_{t-1} \|^2 ]) \\
    & \stackrel{(3e)}{ \leq} (1 - \beta)^2 \mathbb{E}[\|\mathbf{v}_{t-1} - \nabla \mathbf{f}(\mathbf{z_{t-1}}) \|^2] + 2\beta^2 n \bar{\nu}^2 \\
    & + 6(1-\beta)^2L^2 \alpha^2 \mathbb{E}[\|\bar{\mathbf{v}}_{t-1} \|^2] \\
    & + 12(1-\beta)^2L^2\phi_m (n+1) (\mathbb{E}[\mathfrak{L}^2({\z}_t, \phi_t )  + \mathfrak{L}^2({\z}_{t-1}, \phi_{t-1} ) ]), \label{v-nablaf}
\end{aligned}
\end{equation}
where $(3d)$ is due to Cauchy-Schwarz inequality and $(3e)$ is from \eqref{z-x}. 
This completes the proof of the first inequality in Lemma \ref{lemma3}. The upper bound on the averaged version can be derived using the same technique by replacing $\mathbb{E}[\| \mathbf{v}_t - \nabla \mathbf{f}(\mathbf{z_t}) \|^2 ]$ with $\mathbb{E}[\|\bar{ \mathbf{v}}_t - \nabla \bar{\mathbf{f}}(\mathbf{z_t}) \|^2 ]$.
\vspace{-2mm}
\subsection{Proof of Lemma~4}
Since the global function is $L$-smooth, for $\alpha \in (0, \frac{1}{2L}]$ it holds that 

\begin{align*}
   f(\bar{\mathbf{x}}_{t+1}) & \stackrel{(4a)}{ \leq} f(\bar{\mathbf{x}}_{t}) + <\nabla f(\bar{\mathbf{x}}_{t}), \bar{\mathbf{x}}_{t+1} - \bar{\mathbf{x}}_t > + \frac{L}{2} \|\bar{\mathbf{x}}_{t+1} - \bar{\mathbf{x}}_t \|^2 \\
    & \stackrel{(4b)}{ \leq} f(\bar{\mathbf{x}}_{t}) - \alpha <\nabla f(\bar{\mathbf{x}}_{t}),  \bar{\mathbf{v}}_t > +  \frac{L \alpha^2}{2} \|\bar{\mathbf{v}}_{t} \|^2 \\
    & \stackrel{(4c)}{ \leq} f(\bar{\mathbf{x}}_{t}) - \frac{\alpha}{2} \|\nabla f(\bar{\mathbf{x}}_{t}) \|^2 - (\frac{\alpha}{2} - \frac{L\alpha^2}{2}) \|\bar{\mathbf{v}}_{t} \|^2 \\
    & + \frac{\alpha}{2} \|\bar{\mathbf{v}}_t - \nabla f(\bar{\mathbf{x}}_{t})\|^2 \\
    & \stackrel{(4d)}{ \leq} f(\bar{\mathbf{x}}_{t}) - \frac{\alpha}{2} \|\nabla f(\bar{\mathbf{x}}_{t}) \|^2 - (\frac{\alpha}{2} - \frac{L\alpha^2}{2}) \|\bar{\mathbf{v}}_{t} \|^2 \\
    & + \alpha \|\bar{\mathbf{v}}_t - \nabla \bar{\mathbf{f}}(\mathbf{z_t}) \|^2 + \alpha \|\nabla \bar{\mathbf{f}}(\mathbf{z_t}) - \nabla f(\bar{\mathbf{x}}_{t})\|^2 \\
    & \stackrel{(4e)}{ \leq} f(\bar{\mathbf{x}}_{t}) - \frac{\alpha}{2} \|\nabla f(\bar{\mathbf{x}}_{t}) \|^2 - \frac{\alpha}{4} \|\bar{\mathbf{v}}_{t} \|^2 + \alpha \|\bar{\mathbf{v}}_t - \nabla \bar{\mathbf{f}}(\mathbf{z_t}) \|^2 \\
    & + \frac{ \alpha L^2}{n} \|\mathbf{z}_t - \bar{\mathbf{x}}_t \|^2,
\end{align*}
where $(4a)$ is due to $L$-smoothness, $(4b)$ follows from the update of $\x_t$, $(4c)$ is due to the perfect square formula, $(4d)$ is due to the Cauchy-Schwarz inequality and $(4e)$ is due to the smoothness assumption and the range of $\alpha$. Moving $ \|\nabla f(\bar{\mathbf{x}}_{t}) \|^2$ to the left side yields
\begin{align*}
    \|\nabla f(\bar{\mathbf{x}}_{t}) \|^2 & \leq \frac{2 (f(\bar{\mathbf{x}}_{t+1}) - f(\bar{\mathbf{x}}_{t})) }{\alpha } -\frac{1}{2}\|\bar{\mathbf{v}}_{t} \|^2 \\
    & + 2 \|\bar{\mathbf{v}}_t - \nabla \bar{\mathbf{f}}(\mathbf{z_t}) \|^2 +\frac{2L^2}{n}\|\mathbf{z}_t - \bar{\mathbf{x}}_t \|^2.
\end{align*}
By taking the telescoping sum over $t$ from $0$ to $T$,
\begin{equation}
\begin{aligned}
    & \sum_{t=0}^{T-1} \|\nabla f(\bar{\mathbf{x}}_{t}) \|^2 \leq  \frac{2 (f(\bar{\mathbf{x}}_{0}) - f(\bar{\mathbf{x}}_{T})) }{\alpha } -\frac{1}{2}\sum_{t=0}^{T-1}\|\bar{\mathbf{v}}_{t} \|^2 \\
    & + 2\sum_{t=0}^{T-1} \|\bar{\mathbf{v}}_t - \nabla \bar{\mathbf{f}}(\mathbf{z_t}) \|^2 +\frac{2L^2}{n}\sum_{t=0}^{T-1}\|\mathbf{z}_t - \bar{\mathbf{x}}_t \|^2 \\
    & \leq \frac{2 (f(\bar{\mathbf{x}}_{0}) - f(\bar{\mathbf{x}}_{T})) }{\alpha } -\frac{1}{2}\sum_{t=0}^{T-1}\|\bar{\mathbf{v}}_{t} \|^2 + 2\sum_{t=0}^{T-1} \|\bar{\mathbf{v}}_t - \nabla \bar{\mathbf{f}}(\mathbf{z_t}) \|^2 \\ & +\frac{4L^2(n+1)\phi_m}{n}\sum_{t=0}^{T-1}\mathfrak{L}^2({\z}_t, \phi_t ) 
\end{aligned}
\label{eq:34}
\end{equation}
where the last inequality is due to \eqref{z-x}. The proof is completed by taking expectation of both sides of (\ref{eq:34}).

\vspace{-2mm}
\subsection{Proof of Theorem~1}
Using the fact that $\frac{1}{1-(1-\beta)^2} \leq \frac{1}{\beta}$ for $\beta \in (0, 1)$, leveraging the recursive error bounds in inequality $(3e)$ in \eqref{v-nablaf}, and applying telescoping leads to
\begin{align*}
\sum_{t=0}^T \mathbb{E}[\| \mathbf{v}_t - \nabla \mathbf{f}(\mathbf{z_t}) \|^2 ] & \leq \frac{1}{\beta} \mathbb{E}[\| \mathbf{v}_0 - \nabla \mathbf{f}(\mathbf{z_0}) \|^2 ] + 2\beta n \bar{\nu}^2 T \\
& + \frac{1}{\beta} 6(1-\beta)^2L^2 \alpha^2 \sum_{t=1}^T \mathbb{E}[\|\bar{\mathbf{v}}_{t-1} \|^2] \\ 
& + \frac{1}{\beta} 24(1-\beta)^2L^2 \phi_m (n+1) \sum_{t=0}^T \mathbb{E}[\mathfrak{L}^2({\z}_t, \phi_0 ) ].
\end{align*}
Moreover, taking the telescoping sum over Lemma \ref{lemma1} leads to

\begin{align*}
   \sum_{t=0}^T \mathbb{E}\mathfrak{L}^2({\z}_t, \phi_t ) & \leq  \frac{1}{1-(1+\delta^2)/2}  \mathfrak{L}^2({\z}_0, \phi_0 )   \\
   & +  \frac{1}{1-(1+\delta^2)/2}   \frac{2\delta^2 \alpha^2}{1-\delta^2}\sum_{t=0}^T   \mathbb{E}\mathfrak{L}^2({\h}_t, \phi_0 )  \\
    & = \frac{2}{1-\delta^2}  \mathfrak{L}^2({\z}_0, \phi_0 )  + \frac{4\delta^2 \alpha^2}{(1-\delta^2)^2}\sum_{t=0}^T   \mathbb{E}\mathfrak{L}^2({\h}_t, \phi_0 ) ,
\end{align*}
while taking the telescoping sum over Lemma \ref{lemma2} leads to
\begin{align*}
    \sum_{t=0}^T \mathbb{E}[\mathfrak{L}^2({\h}_t, \phi_t ) ]  
    & \leq \frac{2}{1-\delta^2} \mathbb{E}[\|\check{\h}_0 \|^2 ] + \frac{4}{1-\delta^2}\frac{72\delta^2 \alpha^2  L^2 }{(1-\delta^2)^2} \|Y^{-1} \|^2 \sum_{t=0}^{T-1} \mathbb{E}[\|\bar{\mathbf{v}}_{t} \|^2] \\
    &  +  \frac{4\delta^2 }{(1-\delta^2)^2} \|Y^{-1} \|^2 144L^2 \phi_m (n+1) \sum_{t=0}^T \mathbb{E} \mathfrak{L}^2({\z}_t, \phi_t ) \\
    &  + \frac{96\delta^2 }{(1-\delta^2)^2} \|Y^{-1} \|^2 L^2 \beta^2 \sum_{t=0}^T \mathbb{E}[\| \mathbf{v}_t - \nabla \mathbf{f}(\mathbf{z_t}) \|^2 ]\\ 
    & +  \frac{4\delta^2 }{(1-\delta^2)^2} \|Y^{-1} \|^2  24\beta^2  n \bar{\nu}^2 T. 
\end{align*}
Utilizing the upper bounds above, we can derive the range of the step size $\alpha$ such that the upper bound on $\sum_{t=0}^T \mathbb{E}[\mathfrak{L}^2({\h}_t, \phi_t ) ] $ is independent of the other error terms. Letting $\frac{2304\delta^4  \alpha^2}{(1-\delta^2)^4} \|Y^{-1} \|^2 L^2 \phi_m (n+1) +  \frac{9216\delta^4 \alpha^2 }{(1-\delta^2)^4} \|Y^{-1} \|^2 L^4 \phi_m (n+1)  < \frac{1}{2}$, we have
\begin{align*}
    \sum_{t=0}^T \mathbb{E}[\mathfrak{L}^2({\h}_t, \phi_t ) ] & \leq  \frac{4}{1-\delta^2} \mathbb{E}[\mathfrak{L}^2({\h}_0, \phi_0 ) ] + \frac{576\delta^2 \alpha^2  L^2 }{(1-\delta^2)^3} \|Y^{-1} \|^2 \sum_{t=0}^{T-1} \mathbb{E}[\|\bar{\mathbf{v}}_{t} \|^2] \\
    & +  \frac{2304\delta^2 }{(1-\delta^2)^3} \|Y^{-1} \|^2 L^2 (n^2+1) \mathfrak{L}^2({\z}_0, \phi_0 ) \\ 
    & +  \frac{192\delta^2 }{(1-\delta^2)^2} \|Y^{-1} \|^2  \beta^2 n \bar{\nu}^2 T \\
    & + \frac{192\delta^2 }{(1-\delta^2)^2} \|Y^{-1} \|^2 L^2 \beta^2 [\frac{1}{\beta} \mathbb{E}[\| \mathbf{v}_0 - \nabla \mathbf{f}(\mathbf{z_0}) \|^2 ] + 2\beta n \bar{\nu}^2 T \\
& + \frac{1}{\beta} 6(1-\beta)^2L^2 \alpha^2  \sum_{t=1}^T \mathbb{E}[\|\bar{\mathbf{v}}_{t-1} \|^2] \\
&  + \frac{1}{\beta} 24(1-\beta)^2L^2 \phi_m (n+1) \frac{2}{1-\delta^2}  \mathfrak{L}^2({\z}_0, \phi_0 )].
\end{align*}
Initializing by all-zero models and identifying the range of step size $\alpha$ such that the coefficient for the term $\sum_{t=0}^{T-1}\mathbb{E}\|\bar{\mathbf{v}}_{t} \|^2$ is negative completes the proof of Theorem \ref{thm:main}.

\subsection{Proof of Corollary~1.1}
To prove the corollary, we note that
\begin{align*}
    \mathbb{E}[\| \mathbf{v}_0 - \nabla \mathbf{f}(\mathbf{z_0}) \|^2 & = \sum_{i=1}^n \mathbb{E}[ \|\frac{1}{b}\sum_{r=1}^b(\g_{i}(\mathbf{z_0^i}, \xi_{0, r}^i) -\nabla f_i(\mathbf{z_0^i})) \|^2 ] \\
    & = \frac{1}{b^2} \sum_{i=1}^n \sum_{r=1}^b \mathbb{E} [ \| (\g_{i}(\mathbf{z_0^i}, \xi_{0, r}^i) -\nabla f_i(\mathbf{z_0^i})) \|^2 ] \\
    & \leq \frac{n\bar{\nu}^2}{b}
\end{align*}
and
\begin{align*}
    \mathbb{E}[\| \bar{\mathbf{v}}_0 - \nabla \bar{\mathbf{f}}(\mathbf{z_0})  \|^2 & = \mathbb{E}[ \|\frac{1}{n}\sum_{i=1}^n \frac{1}{b}\sum_{r=1}^b(\g_{i}(\mathbf{z_0^i}, \xi_{0, r}^i) -\nabla f_i(\mathbf{z_0^i})) \|^2 ] \\
    & \leq \frac{\bar{\nu}^2}{nb}.
\end{align*}
From the result of Theorem \ref{thm:main}, 
\begin{align*}
    & \frac{1}{T}\sum_{t=0}^{T-1}\mathbb{E} \|\nabla f(\bar{\mathbf{x}}_{t}) \|^2  \leq \frac{2\mathbb{E} (f(\bar{\mathbf{x}}_{0}) - f(\bar{\mathbf{x}}_{T})) }{\alpha T } \\
    & + \frac{2}{\beta T} \mathbb{E}[\|\bar{ \mathbf{v}}_0 - \nabla \bar{\mathbf{f}}(\mathbf{z_0}) \|^2] + 4\beta \bar{\nu}^2 + \mathcal{O}(\frac{\alpha^2}{\beta}(\frac{1}{T}+\beta^2)) \\
    & \leq \frac{2\mathbb{E} (f(\bar{\mathbf{x}}_{0}) - f(\bar{\mathbf{x}}_{T})) }{\alpha T } + \frac{2}{\beta T}\frac{\bar{\nu}^2}{nb} + 4\beta \bar{\nu}^2 + \mathcal{O}(\frac{\alpha^2}{\beta}(\frac{1}{T}+\beta^2)).
\end{align*}
By letting $\alpha = \O(\frac{1}{n^{1/2}T^{1/3}})$, $\beta = \O( \frac{1}{T^{2/3}})$, and $b = \O(\frac{T^{1/3}}{n})$, we conclude that
\begin{equation}
    \frac{1}{T} \sum_{t=0}^{T-1} \mathbb{E}\|\nabla f(\bar{\mathbf{x}}_{t}) \|^2 \leq \O(\frac{1}{T^{2/3}}),
\end{equation}
which completes the proof of the corollary.
\vspace{-2mm}
\subsection{Proof of Lemma~5}
Revisiting the proof of Lemma \ref{lemma4} (specifically, expressions $(4a)-(4e)$) to introduce the PL condition yields

\begin{align*}
    f(\bar{\mathbf{x}}_{t+1}) & \leq f(\bar{\mathbf{x}}_{t}) - <\nabla f(\bar{\mathbf{x}}_{t}), \bar{\mathbf{x}}_{t+1} - \bar{\mathbf{x}}_t > + \frac{L}{2} \|\bar{\mathbf{x}}_{t+1} - \bar{\mathbf{x}}_t \|^2 \\
    & \leq f(\bar{\mathbf{x}}_{t}) - \frac{\alpha}{2} \|\nabla f(\bar{\mathbf{x}}_{t}) \|^2 - \frac{\alpha}{4} \|\bar{\mathbf{v}}_{t} \|^2 + \alpha \|\bar{\mathbf{v}}_t - \nabla \bar{\mathbf{f}}(\mathbf{z_t}) \|^2 \\
    &  + \frac{ \alpha L^2}{n} \|\mathbf{z}_t - \bar{\mathbf{x}}_t \|^2 \\
    & \stackrel{(5a)}{ \leq} f(\bar{\mathbf{x}}_{t}) - \frac{\alpha \mu}{2} ( f(\bar{\mathbf{x}}_{t}) -f^*) - \frac{\alpha}{4} \|\bar{\mathbf{v}}_{t} \|^2 \\
    & + \alpha \|\bar{\mathbf{v}}_t - \nabla \bar{\mathbf{f}}(\mathbf{z_t}) \|^2 + \frac{ \alpha L^2}{n} \|\mathbf{z}_t - \bar{\mathbf{x}}_t \|^2,
\end{align*}
where $(5a)$ is due to the PL condition.
Subtracting $f^* $ from both sides and taking the expectation results in
\begin{align*}
    \mathbb{E}[f(\bar{\mathbf{x}}_{t+1}) - f^*] & \leq \mathbb{E}[ (1-\frac{\alpha \mu}{2}) (f(\bar{\mathbf{x}}_t) - f^* ) - \frac{\alpha}{4} \|\bar{\mathbf{v}}_{t} \|^2 \\
    & + \alpha \|\bar{\mathbf{v}}_t - \nabla \bar{\mathbf{f}}(\mathbf{z_t}) \|^2  + \frac{ \alpha L^2}{n} \|\mathbf{z}_t - \bar{\mathbf{x}}_t \|^2 ]\\
    & \hspace{-0.2in} \leq \mathbb{E}[(1-\frac{\alpha \mu}{2}) (f(\bar{\mathbf{x}}_t) - f^* ) - \frac{\alpha}{4} \|\bar{\mathbf{v}}_{t} \|^2 \\
    & \hspace{-0.2in} + \frac{\alpha}{n} \|\mathbf{v}_t - \nabla \mathbf{f}(\mathbf{z_t}) \|^2 + \frac{ \alpha L^2 \phi_m (2n + 2)}{n}  \mathfrak{L}^2(\z_t, \phi_t) ].
\end{align*}
Having accumulated recursive error bounds for each of the four errors impacting the linear inequality system at iteration $k$, we obtain
\begin{align*}
    \mathbf{u}_{k+1} \leq C \mathbf{u}_k + \mathbf{r}_k.
\end{align*}
\vspace{-2mm}
\subsection{Proof of Lemma~6}
Recall that, as we argued in the discussion following Lemma \ref{lemma5}, the existence of $\x >0$ such that $C \x < \x$ implies $\rho (C) <1$. Building on a similar idea, here we aim to find a range for the step size $\alpha$ and a specific vector $\x >0$ such that $C \x \leq (1 - \frac{\alpha \mu}{4}) \x$, which subsequently implies $\rho (C) \leq 1 - \frac{\alpha \mu}{4} $. To this end, we construct the following inequalities by re-writing $C \x \leq (1 - \frac{\alpha \mu}{4}) \x $ in scalar format for 4-dimensional $\x$ (recall the definition of $C$ in \eqref{eq:c-def}):
\begin{align*}
    & \frac{1+\delta^2}{2} x_1 +  \frac{2\delta^2\alpha^2}{1-\delta^2} x_2 \leq (1 - \frac{\alpha \mu}{4} )x_1,  \\
    & \frac{144\delta^2 \|Y^{-1} \|^2L^2 \phi_m (n + 1)}{1-\delta^2} x_1 + ( \frac{288 \delta^4 \|Y^{-1} \|^2L^2  \phi_m \alpha^2 (n + 1) }{1-\delta^2} \\
    & + \frac{1+\delta^2}{2} ) x_2  + \frac{24\delta^2 \beta^2 \|Y^{-1} \|^2 }{1-\delta^2} x_3  \leq (1 - \frac{\alpha \mu}{4} )x_2, \\
    & 24(1-\beta)^2L^2 \phi_m (n + 1) x_1 +  24(1-\beta)^2L^2 \phi_m (n + 1)\frac{\delta^2\alpha^2}{1-\delta^2} x_2 \\
    & + (1-
    \beta )^2 x_3  \leq (1 - \frac{\alpha \mu}{4} )x_3, \\
    & \frac{2\alpha L^2 \phi_m (n+1)}{n} x_1 + \frac{\alpha}{n} x_3 + (1-\frac{\alpha \mu}{2}) x_4  \stackrel{(6a)}{\leq} (1 - \frac{\alpha \mu}{4} )x_4.
\end{align*}
By rearranging the first three inequalities above, we obtain the following inequalities on $x_1/x_2 $ and $x_3/x_2 $:
\begin{align}
    & \frac{\frac{2\delta^2\alpha^2}{1-\delta^2} }{\frac{1-\delta^2}{2}-\frac{\alpha \mu}{4}}  \leq \frac{x_1}{x_2}, \label{x1x2_1} \\
    & \frac{x_1}{x_2} \leq \frac{\frac{1}{2}(\frac{1-\delta^2}{2}-\frac{\alpha \mu}{4} - \frac{288 \delta^4 \|Y^{-1} \|^2L^2 \phi_m \alpha^2 (n + 1) }{1-\delta^2} ) }{\frac{144\delta^2 \|Y^{-1} \|^2L^2 \phi_m (n + 1)}{1-\delta^2}}, \label{x1x2_2} \\
    & \frac{24(1-\beta)^2L^2 \phi_m(n + 1)\frac{2\delta^2\alpha^2}{1-\delta^2} }{\frac{1}{2}(1 - (1-\beta)^2 - \frac{\alpha \mu}{4})(\frac{1-\delta^2}{2} - \frac{\alpha \mu}{4})}  \leq \frac{x_3}{x_2}, \label{x2x3_1}\\
    & \frac{x_3}{x_2} \leq  \frac{\frac{1}{2}(\frac{1-\delta^2}{2}-\frac{\alpha \mu}{4} - \frac{288 \delta^4 \|Y^{-1} \|^2L^2 \phi_m \alpha^2 (n + 1) }{1-\delta^2} )}{\frac{24\delta^2 \beta^2 \|Y^{-1} \|^2 }{1-\delta^2}}. \label{x2x3_2}
\end{align}
Finally, we set $x_2 = 1$ and solve recursively for $x_1, x_3$ and $x_4$ via $(6a)$ and \eqref{x1x2_1}-\eqref{x2x3_2}. The desired inequality $C \x \leq (1 - \frac{\alpha \mu}{4}) \x $ holds for the following choice of $\x$:
\begin{align*}
    x_1 & = \frac{\frac{2\delta^2\alpha^2}{1-\delta^2} }{\frac{1-\delta^2}{2}-\frac{\alpha \mu}{4}}, \\
    x_2 & = 1, \\
    x_3 & = \frac{96(1-\beta)^2L^2(n^2 +  1)\frac{\delta^2\alpha^2}{1-\delta^2} }{(1 - (1-\beta)^2 - \frac{\alpha \mu}{4})(\frac{1-\delta^2}{2} - \frac{\alpha \mu}{4})}, \\
    x_4 & = \frac{4}{\alpha \mu}[ \frac{2\alpha L^2 \phi_m (n+1)}{n} x_1 + \frac{\alpha}{n} x_3]. 
\end{align*}
Therefore, given the above construction, there exist positive  $x_1, x_2, x_3$ and $x_4$ such that $\rho (C) \leq 1 - \frac{\alpha \mu}{4} $. The corresponding range of $\alpha$ can then be determined via \eqref{x1x2_1}-\eqref{x2x3_2}.
\vspace{-2mm}
\subsection{Proof of Lemma~7}
Recall the definition of $C$ in \eqref{eq:c-def}. To facilitate upcoming analysis of the linear system inequality, let us start by computing the inverse of $I_4 - C$, i.e., 
\[(I_4 - C)^{-1} = \overline{I_4 - C}/\det (I_4 - C),\]
where $\overline{I_4 - C}$ is the adjugate matrix of $I_4 - C$. To compute this inverse, we first find and bound the determinant of $I_4 - C$,
\begin{align*}
    & \det (I_4 - C) \\ & = \frac{\alpha \mu }{2} \{ (\frac{(1-\delta^2)^2}{4} - \frac{144\delta^4 \alpha^2 L^2 \|Y^{-1} \|^2 \phi_m (n+1) }{(1-\delta^2)})(2\beta - \beta^2) \\
    & -24(1-\beta)^2 L^2 \phi_m(n + 1) \frac{48\delta^4 \alpha^2 \beta^2 \|Y^{-1} \|^2 }{(1-\delta^2)^2} \\
    & - \frac{576\delta^4 \alpha^2 \|Y^{-1} \|^2 L^2 \phi_m(n + 1) (1-(1 - \beta)^2 )}{(1-\delta^2)^2} \\
    & - 288(1-\beta)^2L^2\phi_m (n + 1)\delta^4 \alpha^2 \beta^2 \frac{ \|Y^{-1} \|^2}{1-\delta^2}
    \} \\
    & \stackrel{(7a)}{\geq} \frac{\alpha \mu}{6}   (\frac{(1-\delta^2)^2}{4} - \frac{144\delta^4 \alpha^2 L^2 \|Y^{-1} \|^2 \phi_m (n+1) }{(1-\delta^2)})(2\beta - \beta^2). 
\end{align*}
The lower bound $(7a)$ is achieved if the learning rate/stepsize parameters $\alpha$ and $\beta$ are chosen such that
\begin{align*}
     & 24(1-\beta)^2 L^2\phi_m(n + 1) \frac{48\delta^4 \alpha^2 \beta^2 \|Y^{-1} \|^2 }{(1-\delta^2)^2} \\
     & + \frac{576\delta^4 \alpha^2 \|Y^{-1} \|^2 L^2 \phi_m(n + 1) (2\beta - \beta^2 )}{(1-\delta^2)^2} \\ & + 288(1-\beta)^2L^2 \phi_m (n + 1)\delta^4 \alpha^2 \beta^2 \frac{ \|Y^{-1} \|^2}{1-\delta^2} \\
     & 
     \leq    (\frac{(1-\delta^2)^2}{6} - \frac{96\delta^4 \alpha^2 L^2 \|Y^{-1} \|^2 \phi_m (n+1) }{1-\delta^2})(2\beta - \beta^2).
\end{align*}
\textcolor{black}{This also implies that 
\begin{align*}
    \frac{288\delta^4 \alpha^2 L^2 \|Y^{-1} \|^2 \phi_m (n+1) }{(1-\delta^2)} \leq \frac{(1-\delta^2)^2}{8}.
\end{align*}
By invoking $(7a)$},
we can show that $\det (I_4 - C) \geq \frac{\alpha \mu \beta (1-\delta^2)^2}{96} $.
Next, we note that the adjugate matrix needed to specify the inverse of $I_4 - C$ has entries that satisfy the following (in)equalities:
\begin{align*}
    [\overline{I_4 - C}]_{1, 2} & = \frac{2 \alpha^3 \beta \mu \delta^2}{1-\delta^2} - \frac{\alpha^3 \beta^2 \mu \delta^2}{1-\delta^2} \\
    [\overline{I_4 - C}]_{1, 3} & = \frac{24 \| Y^{-1}\|^2 \alpha^3 \beta^2 \mu \delta^4 }{(1-\delta^2)^2} \\
    [\overline{I_4 - C}]_{2, 2} & = \frac{\alpha \mu \beta (1-\delta^2)}{2} - \frac{\alpha \mu \beta^2 (1-\delta^2)}{4} \\
    [\overline{I_4 - C}]_{2, 3} & = 6 \|Y^{-1} \|^2 \alpha \beta^2 \mu \delta^2 \\
    [\overline{I_4 - C}]_{3, 2} & \leq \frac{30 L^2 \alpha^3 \mu \delta^2 \phi_m(1 + \beta^2  + n   + \beta^2 n  )}{1-\delta^2} \\
    [\overline{I_4 - C}]_{3, 3} & \leq \frac{\alpha \mu}{8} \\
    [\overline{I_4 - C}]_{4, 2} & \leq \frac{ L^2 \alpha^3 \delta^2 \phi_m (60 + 56\beta^2 + 60n + 56 \beta^2 n )}{n(1-\delta^2)} \\
    [\overline{I_4 - C}]_{4, 3} & \leq \frac{\alpha(1-\delta^2)^2}{4n} \\ & + \frac{L^2 \|Y^{-1} \|^2 \alpha^3 \delta^4 \phi_m (96\beta^2 + 144 \delta^2 + 96\beta^2 n )}{n(1-\delta^2)^2}. 
\end{align*}
Recall \eqref{eq:linear_system} 
which states that for all $k \in [1, T]$,
\begin{align*}
    \mathbf{u}_k \leq C^k \mathbf{u}_0 + \sum_{t = 0}^{k-1}C^t \mathbf{r}_k \leq C^k \mathbf{u}_0 + \sum_{t = 0}^{k-1}C^t \mathbf{r}_{k_m} \\ \leq C^k \mathbf{u}_0 + (I_4 - C)^{-1} \mathbf{r}_{k_m},
\end{align*}
where $\mathbf{r}_{k_m} = \max_{k \in [1, T]} \mathbf{r}_k$ and $k_m = \mathrm{argmax}_{k \in [1, T] } \mathbf{r}_k$.

Since $\mathbf{r}_k $ is a function of $ E[\|\bar{\mathbf{v}}_{k} \|^2]$,
we proceed by deriving an upper bound on $E[\|\bar{\mathbf{v}}_{k} \|^2] $. Following the proof of Lemma \ref{lemma5},
\begin{align*}
    \frac{\alpha}{4} \mathbb{E}[\|\bar{\mathbf{v}}_{k} \|^2] & \leq \mathbb{E} [(1-\frac{\alpha \mu}{2}) (f(\bar{\mathbf{x}}_k) - f^* ) \\
    & + \frac{\alpha}{n} \|\mathbf{v}_k - \nabla \mathbf{f}(\mathbf{z_k}) \|^2  + \frac{ \alpha L^2 \phi_m (2n + 2)}{n}  \mathfrak{L}^2(\z_t, \phi_t)] \\
    & \stackrel{(7b)}{\leq} C_{exp} + (1-\frac{\alpha \mu}{2})[(I_4 - C)^{-1} \mathbf{r}_k]_4 \\
    & + \frac{\alpha}{n} [(I_4 - C)^{-1} \mathbf{r}_k]_3 +\frac{ \alpha L^2 \phi_m (2n + 2)}{n} [(I_4 - C)^{-1} \mathbf{r}_k]_1,
\end{align*}
where
\begin{align*}
    C_{exp} & = \max_{k_m} \{ (1-\frac{\alpha \mu}{2})[C^{k_m} \mathbf{u}_0]_4 + \frac{\alpha}{n}[C^{k_m} \mathbf{u}_0]_3 \\
    & + \frac{\alpha L^2\phi_m (2n + 2)}{n}[C^{k_m} \mathbf{u}_0]_1 \}.
\end{align*}
We leverage the error bounds in Lemmas~\ref{lemma1}, \ref{lemma2}, \ref{lemma3} and \ref{lemma5} and the computation of $(I_4 - C)^{-1} $  in $(7b)$ to establish
\begin{align*}
    \frac{1}{4} \mathbb{E}[\|\bar{\mathbf{v}}_{k} \|^2] & \leq \frac{C_{exp}}{\alpha} + \frac{96(1-\frac{\alpha \mu}{2})}{ \mu \beta (1-\delta^2)^2}  [\frac{ L^2 \alpha \delta^2 \phi_m (60 + 56\beta^2 + 60n + 56 \beta^2 n )}{n(1-\delta^2)}  \\
    & \frac{8\delta^2}{1-\delta^2} \|Y^{-1} \|^2 (9\alpha^2 L^2 \mathbb{E}[\|\bar{\mathbf{v}}_{k} \|^2] + 3\beta^2 \bar{\nu}^2 n ) \\
    & + (\frac{(1-\delta^2)^2}{4\alpha n} + \frac{L^2 \|Y^{-1} \|^2 \alpha \delta^4 \phi_m (96\beta^2 + 144\delta^2 + 96\beta^2 n )}{n(1-\delta^2)^2}) \\
    & (6(1-\beta)^2L^2 \alpha^2 \mathbb{E}[\|\bar{\mathbf{v}}_{k} \|^2] + 2\beta^2\bar{\nu}^2n ) ] \\
    & + \frac{1}{n} \frac{96}{\beta (1-\delta^2)^2} \times [ \frac{30 L^2 \alpha^2 \delta^2\phi_m (1 + \beta^2  + n  + \beta^2 n )}{1-\delta^2} \frac{8\delta^2}{1-\delta^2} \\
    & \|Y^{-1} \|^2 (9\alpha^2 L^2 \mathbb{E}[\|\bar{\mathbf{v}}_{k} \|^2] + 3\beta^2 \bar{\nu}^2 n ) \\
    & + \frac{1}{8} (6(1-\beta)^2L^2 \alpha^2 \mathbb{E}[\|\bar{\mathbf{v}}_{k} \|^2] + 2\beta^2\bar{\nu}^2n )  ] \\
    & +\frac{ L^2 \phi_m (2n + 2)}{n} \frac{96}{ (1-\delta^2)^2} \times [(\frac{2 \alpha^2  \delta^2}{1-\delta^2} - \frac{\alpha^2 \beta  \delta^2}{1-\delta^2})\frac{8\delta^2}{1-\delta^2} \|Y^{-1} \|^2 \\
    & (9\alpha^2 L^2 \mathbb{E}[\|\bar{\mathbf{v}}_{k} \|^2] + 3\beta^2 \bar{\nu}^2 n ) \\
    & + \frac{24 \| Y^{-1}\|^2 \alpha^2 \beta \delta^4 }{(1-\delta^2)^2}(
     6(1-\beta)^2L^2 \alpha^2 \mathbb{E}[\|\bar{\mathbf{v}}_{k} \|^2] + 2\beta^2\bar{\nu}^2n )].
\end{align*}
Collecting the terms with $\mathbb{E}[\|\bar{\mathbf{v}}_{k} \|^2] $ on the right-hand side and imposing 
\begin{align*}
    & (1-\frac{\alpha \mu}{2})\frac{96}{ \mu \beta (1-\delta^2)^2} \times \\
    & [\frac{ 9 L^4 \alpha^3 \delta^2 \phi_m (60 + 56\beta^2 + 60n + 56 \beta^2 n )}{(1-\delta^2)} \frac{8\delta^2}{1-\delta^2} \|Y^{-1} \|^2  \\
    & + (\frac{(1-\delta^2)^2}{4\alpha n} + \frac{L^2 \|Y^{-1} \|^2 \alpha \delta^4 \phi_m (96\beta^2 + 144 \delta^2 + 96\beta^2 n )}{n(1-\delta^2)^2}) \\
    & (6(1-\beta)^2L^2 \alpha^2 )] \\
    & +  \frac{96}{n\beta (1-\delta^2)^2} [ \frac{2160 L^4 \alpha^4 \delta^4\phi_m (1 + \beta^2  + n  + \beta^2 n )}{(1-\delta^2)^2}  \|Y^{-1} \|^2  \\
    & +  \frac{3}{4} (1-\beta)^2L^2 \alpha^2] \\
    & + \frac{96 L^2 \phi_m (2n + 2)}{n(1-\delta^2)^2}  [(\frac{8 \alpha^2  \delta^2}{1-\delta^2} - \frac{\alpha^2 \beta  \delta^2}{1-\delta^2})\frac{2\delta^2}{1-\delta^2} \|Y^{-1} \|^2 \times 9\alpha^2 L^2 \\
    & + \frac{144 \| Y^{-1}\|^2 \alpha^2 \beta \delta^4 }{(1-\delta^2)^2}
     (1-\beta)^2L^2 \alpha^2 ]  \leq \frac{1}{8}
\end{align*}
leads to
\begin{equation}\label{eq:v_k}
\begin{aligned}
    \frac{1}{8} \mathbb{E}[\|\bar{\mathbf{v}}_{k} \|^2] & \leq \frac{C_{exp}}{\alpha} + \frac{96(1-\frac{\alpha \mu}{2})}{ \mu  (1-\delta^2)^2} [\frac{8 L^2 \alpha \delta^4 \phi_m (60 + 56\beta^2 + 60n + 56 \beta n )}{n(1-\delta^2)^2} \\
    &  \|Y^{-1} \|^2  3\beta \bar{\nu}^2 n  + (\frac{(1-\delta^2)^2}{4\alpha } \\
    & + \frac{L^2 \|Y^{-1} \|^2 \alpha \delta^4 \phi_m (96\beta^2 + 144 \delta^2 + 96\beta^2 n )}{(1-\delta^2)^2}) 2\beta^2\bar{\nu}^2  ] \\
    & + \frac{1}{n} \frac{96}{ (1-\delta^2)^2} \times [ \frac{30 L^2 \alpha^2 \delta^2 \phi_m (1 + \beta^2  + n  + \beta^2 n )}{1-\delta^2} \frac{8\delta^2}{1-\delta^2} \\
    & \|Y^{-1} \|^2  3\beta \bar{\nu}^2 n  + \frac{2}{8}\beta \bar{\nu}^2n   ] \\
    & +\frac{ L^2 \phi_m (2n + 2)}{n} \frac{96}{ (1-\delta^2)^2} \times [(\frac{2 \alpha^2  \delta^2}{1-\delta^2} - \frac{\alpha^2 \beta  \delta^2}{1-\delta^2})\frac{8\delta^2}{1-\delta^2} \\
    & \|Y^{-1} \|^2  3\beta^2 \bar{\nu}^2 n  + \frac{48 \| Y^{-1}\|^2 \alpha^2 \beta \delta^4 }{(1-\delta^2)^2} \beta^2\bar{\nu}^2n ].
\end{aligned}
\end{equation}
Since the right hand side is independent of $k$ and $T$, this expression provides an upper bound on $E[\|\bar{\mathbf{v}}_{k} \|^2] $ w.r.t. $\bar{\nu}$ for all $k, T$.
\vspace{-2mm}
\subsection{Proof of Theorem~2}
We recall conditions on the step sizes $\alpha$ and $\beta$ used in Lemmas~\ref{lemma1}-\ref{lemma3} and \ref{lemma5}-\ref{lemma7}, and let the step sizes be such that
\begin{align*}
    & \alpha \leq \min\{\frac{1}{2L}, \; \frac{2\beta}{\mu}, \; \frac{1-\delta^2}{\mu}\}, \\
    & 576 L^2\phi_m (n + 1)\frac{\delta^4\alpha^2 \|Y^{-1} \|^2 }{(1-\delta^2)^2} \leq   \frac{(1-\delta^2)\alpha \mu}{32} \times \\
    & \times (\frac{1-\delta^2}{4} - \frac{288 \delta^4 \|Y^{-1} \|^2L^2 \alpha^2 \phi_m (n + 1) }{1-\delta^2} ), \\
    & 1728 L^2\phi_m (n + 1) \frac{\delta^4 \alpha^2 \|Y^{-1} \|^2 }{(1-\delta^2)^2}   + 384 L^2 \phi_m(n + 1)\delta^4 \alpha^2 \frac{ \|Y^{-1} \|^2}{1-\delta^2}
    \\ & \leq    \frac{(1-\delta^2)^2}{6}, \\
    & \mbox{ and } \\    
     & \frac{96}{ \mu }  [\frac{3 L^4 \alpha \delta^4 \phi_m (356 + 212n )\|Y^{-1} \|^2 }{(1-\delta^2)^4}  + \frac{3L^2 \alpha }{2 }  ] \\ & + 96  [ \frac{2736 L^4 \alpha^2 \delta^4 \phi_m ( n + 1 ) \|Y^{-1} \|^2}{(1-\delta^2)^4}  +  \frac{3L^2 \alpha^2 }{4(1-\delta^2)^2}]  \leq \frac{\beta}{8}.
\end{align*}

\noindent
Then, it holds that
\begin{align*}
    \lim_{k \to \infty} \sup \mathbb{E}[f(\bar{\mathbf{x}}_{k+1}) - f^*] & \leq \frac{96}{ \mu \beta (1-\delta^2)^2}  [\frac{8L^2 \alpha^2 \delta^4\|Y^{-1} \|^2 \phi_m (60 + 56\beta^2 + 60n + 56 \beta^2 n )}{(1-\delta^2)^2}   \\ & (9\alpha^2 L^2 \mathbb{E}[\|\bar{\mathbf{v}}_{k} \|^2] + 3\beta^2 \bar{\nu}^2 ) \\
    & + (\frac{(1-\delta^2)^2}{4} + \frac{L^2 \|Y^{-1} \|^2 \alpha^2 \delta^4 \phi_m (96\beta^2 + 144 \delta^2 + 96\beta^2 n )}{(1-\delta^2)^2}) \\ & (6(1-\beta)^2L^2 \alpha^2  \mathbb{E}[\|\bar{\mathbf{v}}_{k} \|^2] + 2\beta^2\bar{\nu}^2 ) ] \\
    & \leq (\frac{\alpha^2 L^2 }{4} + \frac{3 L^2 \alpha^2}{2\beta } ) 
    \mathbb{E}[\|\bar{\mathbf{v}}_{k} \|^2] + \frac{96}{ \mu \beta (1-\delta^2)^2} \\ & [\frac{8 L^2 \alpha^2 \delta^4 \phi_m (60 + 56\beta^2 + 60n + 56 \beta^2 n )}{(1-\delta^2)^2}  \|Y^{-1} \|^2 ( 3\beta^2 \bar{\nu}^2 ) + \\ & \frac{(1-\delta^2)^2}{4} ( 2\beta^2\bar{\nu}^2 )+\\ & \frac{L^2 \|Y^{-1} \|^2 \alpha^2 \delta^4 \phi_m (96\beta^2 + 144 \delta^2 + 96\beta^2 n )}{(1-\delta^2)^2} ( 2\beta^2\bar{\nu}^2 )].
\end{align*}
Substituting the upper bound derived for $\mathbb{E}[\|\bar{\mathbf{v}}_{k} \|^2]$ in \eqref{eq:v_k}, one can readily show that $\mathbb{E}[f(\bar{\mathbf{x}}_{k+1}) - f^*] $ decays linearly to the steady-state error given in the statement of Theorem \ref{thm:main2}.


\clearpage
\bibliographystyle{acm}
\bibliography{n_bib.bib}

\end{document}